\documentclass{article}

\usepackage{arxiv}

\usepackage[utf8]{inputenc} % allow utf-8 input
\usepackage[T1]{fontenc}    % use 8-bit T1 fonts
\usepackage{hyperref}       % hyperlinks
\usepackage{url}            % simple URL typesetting
\usepackage{booktabs}       % professional-quality tables
\usepackage{amsfonts}       % blackboard math symbols
\usepackage{nicefrac}       % compact symbols for 1/2, etc.
\usepackage{microtype}      % microtypography
\usepackage{lipsum}		% Can be removed after putting your text content
\usepackage{graphicx}
\usepackage[numbers]{natbib}
\usepackage{doi}
\usepackage{xcolor}
\usepackage{color, colortbl}
\usepackage{subcaption}
\usepackage{graphicx}
\usepackage{multirow}
\usepackage{longtable}

\title{From Reflection to Repair: A Scoping Review of Dataset Documentation Tools}

\author{ \href{https://orcid.org/0000-0001-5690-1850}{\includegraphics[scale=0.06]{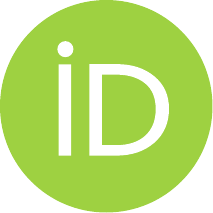}\hspace{1mm}Pedro Reynolds-Cuéllar}\thanks{Authors contributed equally to the paper.} \\
	Robotics and AI Institute\\
	Cambridge, MA 02142 \\
	\texttt{pcuellar@rai-inst.com} \\
	\And
	\href{https://orcid.org/0000-0002-8071-6716}{\includegraphics[scale=0.06]{figures/orcid.pdf}\hspace{1mm}Marisol Wong-Villacres}$^*$ \\
	Escuela Superior Politécnica del Litoral\\
	Guayaquil - Ecuador \\
	\texttt{lvillacr@espol.edu.ec} \\
	\And
	\href{https://orcid.org/0000-0002-4230-3777}{\includegraphics[scale=0.06]{figures/orcid.pdf}\hspace{1mm}Adriana Alvarado Garcia} \\
	IBM Research\\
    USA \\
	\texttt{adriana.ag@ibm.com} \\
	\And
	\href{https://orcid.org/0009-0005-2580-7040}{\includegraphics[scale=0.06]{figures/orcid.pdf}\hspace{1mm}Heila Precel} \\
	Robotics and AI Institute\\
	Cambridge, MA 02142 \\
	\texttt{hprecel@rai-inst.com} \\
}

\renewcommand{\shorttitle}{From Reflection to Repair: A Scoping Review of Dataset Documentation Tools}

\hypersetup{
pdftitle={From Reflection to Repair: A Scoping Review of Dataset Documentation Tools},
pdfsubject={cs.CY, cs.RO, cs.HC},
pdfauthor={Pedro Reynolds-Cuéllar, Marisol Wong-Villacres, Adriana Alvarado Garcia, Heila Precel},
pdfkeywords={Dataset, Documentation, HCI, Data Work, Data Practices},
}

\begin{document}
\maketitle

\begin{abstract}
	Dataset documentation is widely recognized as essential for the responsible development of automated systems. Despite growing efforts to support documentation through different kinds of artifacts, little is known about the motivations shaping documentation tool design or the factors hindering their adoption. We present a systematic review supported by mixed-methods analysis of 59 dataset documentation publications to examine the motivations behind building documentation tools, how authors conceptualize documentation practices, and how these tools connect to existing systems, regulations, and cultural norms. Our analysis shows four persistent patterns in dataset documentation conceptualization that potentially impede adoption and standardization: unclear operationalizations of documentation’s value, decontextualized designs, unaddressed labor demands, and a tendency to treat integration as future work. Building on these findings, we propose a shift in Responsible AI tool design toward institutional rather than individual solutions, and outline actions the HCI community can take to enable sustainable documentation practices. 
\end{abstract}

% keywords can be removed

\section{Introduction}
Increasingly, Human-Computer Interaction (HCI) research in Responsible AI (RAI) has stressed that supporting the careful interrogation and understanding of datasets can facilitate the ethical construction of automated predictive technologies across high-stakes domains (e.g., healthcare, criminal justice, employment, education). To help data practitioners engage with much needed ethical data management practices, Responsible AI and HCI scholars---among others---have increasingly championed data documentation as a preferred mechanism \cite{gebru_datasheets_2021,chmielinskiDatasetNutritionLabel2022a,NIST_2023}. One of the main reasons is the traceability that documentation provides in interpreting decisions made by automated systems. A rapid rise in tools to support this practice has followed, including frameworks, toolkits, and applications. Studies exploring practitioners’ perspectives on documentation, however, suggest that despite the rapid growth of these tools, standardization and adoption of documentation practices remains stagnant \cite{Bhardwaj_Gujral_Wu_Zogheib_Maharaj_Becker_2024,heger_understanding_2022}. Scholars such as~\citeauthor{hutchinson_towards_2021} argue that misunderstanding dataset development cultures and overlooking AI dataset expertise are among the factors at play in this phenomenon~\cite{hutchinson_towards_2021}. However, the role of dataset documentation tools in feeding into adoption barriers remains unexplored. 

In this paper, we use this gap as an opportunity to explore how creators conceptualize documentation tools, and how those conceptualizations might help or hinder mainstream adoption and standardization. Specifically, our work presents a systematic exploration of a corpus of fifty-nine academic contributions across several computing, machine learning, and Responsible AI venues. Each contribution discusses the use of tools or frameworks for dataset documentation. We investigate three aspects of documentation tools that could critically shape users’ adoption: first, the goals and motivations behind tool creation; second, how tool creators understand the dataset documentation process itself; and third, these tools’ efforts to connect with larger systems, including regulatory frameworks. Our findings highlight how creators often motivate their work via transparency and accountability, but take radically different routes towards these goals based on their own definition of what productive documentation practices entail. Further, creators tend to offer overly general tools that stem from their own needs rather than those of users. As a result, all documentation tools, even those that rely on automation to minimize implementation effort, end up imposing diverse forms of labor on individuals while overlooking the organizational, collective, and infrastructural support required to sustain effective practices. Finally, our findings highlight that without offering concrete guidance and evidence on how to integrate these tools, integration efforts are mostly aspirational. 

Our work makes three primary contributions to the literature. First, we provide a scoping review of the dataset documentation landscape, analyzing tools to identify gaps and common threads to help inform adoption efforts. Second, we offer a mixed-method characterization of how the diversity of goals, conceptualizations, and integration efforts behind these tools fragments the value of documenting datasets, critically impacting their adoption and standardization. Lastly, we discuss how documentation fragmentation stems from a systemic gap between Responsible AI research---specifically around tools---and practice, and outline a research agenda for HCI research to support an institutional---rather than individual---design of documentation tools. Together, these contributions aim to advance the design and integration of Responsible AI tools.

\section{Related Work}
\label{sec:related-work}

\subsection{Responsible AI Tool Design in HCI}
\label{sec:related:rai-tool-design}
The increasing prominence of LLMs and other automated systems has brought about both massive transformation and disruption to our technological ecosystems~\cite{manduchi2024challenges}. With it, concerns about datasets used in model development and evaluation have vaulted to the forefront of public discourse~\cite{Roe_Perkins_2023}. Critical scholarship across ML and HCI has demonstrated that datasets can be prone to issues across numerous axes~\cite{Bender_Gebru_McMillan-Major_Shmitchell_2021, Paullada_Raji_Bender_Denton_Hanna_2021}, including the over-or-under-representation of subjects across protected characteristics~\cite{Buolamwini_2017,Bolukbasi_Chang_Zou_Saligrama_Kalai_2016,Sap_Card_Gabriel_Choi_Smith_2019,Shankar_Halpern_Breck_Atwood_Wilson_Sculley_2017}; difficult-to-remove hate speech, sexually exploitative material, personally identifiable information, and intellectual property~\cite{Luccioni_Viviano_2021,Thiel_2023,Hong_Hutson_Agnew_Huda_Kohno_Morgenstern_2025,Longpre_Mahari_Chen_Obeng-Marnu_Sileo_Brannon_Muennighoff_Khazam_Kabbara_Perisetla_et_al._2024,Precel_McDonald_Hecht_Vincent_2024}; and poor working conditions and mental health issues for data labelers, to mention a few~\cite{Gray_Suri_2019,Gillespie_2019}. Once data is used to train automated systems, other detrimental downstream effects come into view, including allocation and representational harms~\cite{Abbasi_Friedler_Scheidegger_Venkatasubramanian_2019,Crawford_2017}, hostile model responses that are resistant to finetuning~\cite{Qi_Zeng_Xie_Chen_Jia_Mittal_Henderson_2023}, hallucinations and security vulnerabilities~\cite{Dziri_Milton_Yu_Zaiane_Reddy_2022, Vassilev_2025}, and labor displacement~\cite{Eloundou_Manning_Mishkin_Rock_2023}.

The HCI community has responded by providing insight into how practitioners engage and interact with data, as well as by advancing approaches to designing adequate RAI tools. In addition to paving the way on the critical analysis of automated systems and providing a theoretical backing for analyzing resulting harms~\cite{Orr_Crawford_2024, Paullada_Raji_Bender_Denton_Hanna_2021,Crawford_2017, Abbasi_Friedler_Scheidegger_Venkatasubramanian_2019}, HCI researchers have also advanced user studies about the design and efficacy of Responsible AI frameworks and toolkits in practice~\cite{Holstein_Vaughan_III_Dudík_Wallach_2019, Madaio_Chen_Wallach_Wortman_Vaughan_2024, Balayn_Yurrita_Yang_Gadiraju_2023, Deng_Nagireddy_Lee_Singh_Wu_Holstein_Zhu_2022, Lee_Singh_2021,Shen_Wang_Deng_Brusse_Velgersdijk_Zhu_2022}. For example, findings point out that while toolkits are helpful for identifying and mitigating fairness issues, they also have limitations.~\citeauthor{Balayn_Yurrita_Yang_Gadiraju_2023} note that toolkits \textit{``create a gateway to a narrow view on distributive justice''}~\cite{Balayn_Yurrita_Yang_Gadiraju_2023}, shrinking the perceived problems and solutions to those supported by the tool. 
These kinds of studies help the community interrogate documentation tools and establish ways in which they can be re-designed or adopted. Different kinds of tools across multiple stages of automated system development have been proposed, including model documentation frameworks, fairness and interpretability toolkits, and educational materials~\cite{Prem_2023}. These tools cover a variety of processes, including e.g., data curation and bias detection~\cite{cinca_practitioners_2025,kabir_stile_2024}. Studies report that while these tools can help practitioners navigate questions around RAI, they often require contextual guidance and can lead to steep learning curves making their adoption and integration difficult~\cite{Lee_Singh_2021}.

\subsection{Dataset Documentation as a Solution for Responsible AI}
\label{sec:related:dataset-documentation-helps}
In recent years, dataset documentation has emerged as a salient complement to fairness toolkits and model design frameworks, but with a specific focus on dataset-driven AI harms. Researchers have argued that ``documentation has emerged as 
an essential component of AI transparency and a foundation for Responsible AI development''~\cite{The_CLeAR_Documentation_Framework_for_AI_Transparency_2024}, and the World Economic Forum has recommended that all entities ``develop standards to track the provenance, development, and use of training data sets throughout their life cycle''~\cite{Global_Future_Council_on_Human_Rights_2018}. Specifically, as suggested by the computing truism ``garbage in, garbage out'', metadata that describes a dataset's sourcing, provenance, and precise contents can enable model developers to assess and mitigate bias down the line. Maintaining thorough documentation makes it possible to assess and resolve IP-related harms or develop strategies for compensating data creators when their data is used by a model contingent on identifying original creators~\cite{Sarcevic_Karlowicz_Mayer_Baeza-Yates_Rauber_2024,Longpre_Mahari_Chen_Obeng-Marnu_Sileo_Brannon_Muennighoff_Khazam_Kabbara_Perisetla_et_al._2024,Kandpal_Raffel_2025}. Researchers have also argued that the reflection triggered by the documentation process is itself a forcing function for dataset creators to predict and stem potential sources of bias~\cite{gebru_datasheets_2021}. 

In the United States, the push for dataset documentation sits within a fragmented and developing regulatory landscape. At the federal level, two consecutive administrations have incorporated data transparency into their AI priorities; the now-repealed Executive Order 14110 calls for ``authenticating content and tracking its provenance''~\cite{president2023safe} while America's AI Action Plan emphasizes increasing data quality standards across the board and developing open-source, open-weight, and interpretable models and datasets~\cite{Trump_AI_Action_Plan_2025}. In the National Institute of Standards and Technology (NIST)’s guide to AI risk management, dataset documentation is highlighted as a key strategic element for addressing security, transparency, and accountability concerns~\cite{NIST_2023}. However, despite identifying documentation as a priority, there are neither concrete federal guidelines about the right frameworks to follow nor are there oversight mechanisms to ensure compliance.

In response, states have started to implement their own AI-related legislation. California's Generative AI training data transparency act~\cite{AB-2013_California_Gen_AI}, for example, requires that upon model release, all AI developers also release documentation describing the general composition, provenance, and processing of all training data used. Colorado's AI Act~\cite{SB24-205_Colorado_Act} identifies Data Cards~\cite{Data_Cards_Pushkarna_Zaldivar_Kjartansson_2022} specifically, requiring them to be released alongside high-risk models. Meanwhile, intellectual property issues are currently being litigated via numerous court cases representing the interests of---among others---artists, writers, journalists, and musicians~\cite{AndersenStabilityAI,ConcordMusicGroup2024,DowJonesCompany,ReOpenAIInc}. Without clarity around national requirements, a number of independent organizations have proposed their own governance solutions. Key efforts here include Stanford University's Foundation Model Transparency index, which independently assesses publicly available models on transparency (including of training data),~\cite{Bommasani_Klyman_Kapoor_Longpre_Xiong_Maslej_Liang} and Behavioral Use Licensing, proposing a voluntary licensing model for datasets to enforce stated use cases~\cite{Contractor_McDuff_Haines_Lee_Hines_Hecht_Vincent_Li_2022}. Academic institutions and some industrial products have also made efforts to self-regulate: NeurIPS’ ethics requirements~\cite{NeurIPSCodeEthics}, for example, require data and model documentation as part of submission, and Hugging Face’s Data Cards~\cite{HuggingFaceDataset} allow for documentation to any datasets published via the site.

\subsection{Design and Adoption of Documentation Tools}
\label{sec:related:governance-solutions}
As dataset documentation garnered interest across academic, corporate, and regulatory spheres, researchers in the HCI community have proposed numerous general-purpose documentation tools, understood as structured artifacts (e.g., software applications, batteries of questions, schemas) that facilitate the recording of key information across different stages of a dataset's lifecycle. Examples include, but are by not limited to, Datasheets for Datasets~\cite{gebru_datasheets_2021}, Data Cards~\cite{Data_Cards_Pushkarna_Zaldivar_Kjartansson_2022}, the Data Requirements section of Microsoft's RAI Impact Assessments~\cite{Microsoft_2022}, Aether's Dataset Documentation Template~\cite{Microsoft_2022_Aether}, FactSheets~\cite{Fact_Sheets_Arnold_Bellamy_Hind_Houde_Mehta_Mojsilović_Nair_Ramamurthy_Olteanu_Piorkowski_2019}, Dataset Nutrition Labels~\cite{holland_dataset_2018}, and Data Statements~\cite{bender_data_2018, mcmillan-major_data_2023}.

These frameworks generally take the form of structured questions for dataset creators to answer before and during dataset development. In addition to providing downstream users with information about data quality, provenance, and known issues, they are typically also designed ``to encourage careful reflection on the process of creating, distributing, and maintaining a dataset, including any underlying assumptions, potential risks or harms, and implications of use''~\cite{gebru_datasheets_2021}. These efforts reflect a desire across industry and academia for mechanisms to foreground key information associated with dataset production and further use. However, as compared to the fairness tooling ecosystem we discussed in Section \ref{sec:related:rai-tool-design}, not much research has directly investigated the motivation, design choices, or subsequent (lack of) adoption of documentation tools by stakeholders. An empirical basis for understanding both how these tools come to be---and what elements contribute to their adoption---is crucial to inform the HCI community's next steps towards effective dataset documentation.

Existing research suggests that dataset documentation is difficult to achieve in practice. First, datasets are difficult if not impossible to retract once published, and formal retraction is limited for preventing continued distribution~\cite{Peng_Mathur_Narayanan_2021}. In one dramatic but not atypical example described by~\citeauthor{Schneider_Ye_Hill_Whitehorn_2020}, falsified clinical trial data ``continues to be cited positively and uncritically... eleven years after its retraction''~\cite{Schneider_Ye_Hill_Whitehorn_2020}. Documenting data after release can be just as difficult, with potential issues ranging from multiple competing versions of a single dataset to unrecorded procurement practices~\cite{Bandy_Vincent_2021}. Meanwhile, web-scale data collected via internet crawls can include petabytes of data; one popular dataset, the Common Crawl, includes text from over 1 billion pages~\cite{Common_Crawl}. Documenting each page is infeasibly resource-intensive.

Even when technically possible, diligent data documentation is rarely aligned with engineers’ interests.~\citeauthor{Orr_Crawford_2024} highlight ``the messy and contingent realities of dataset preparation,'' with a focus on four competing elements: dataset scale, limited access to resources, reliance on shortcuts, and ambivalence regarding accountability for the final product~\cite{Orr_Crawford_2024}.~\citeauthor{Alvarado_Garcia_Candello_Badillo-Urquiola_Wong-Villacres_2025} found that ``informed by LLM's need for scale and nascent community-based recommendations, practitioners...[made] decisions that primarily ensured scale and cost-effectiveness,'' with the slow process of developing high quality documentation falling by the wayside.~\citeauthor{Bhardwaj_Gujral_Wu_Zogheib_Maharaj_Becker_2024} and~\citeauthor{Yang_Liang_Zou_2023} conducted quantitative analyses investigating ``dataset cards'' on Hugging Face~\cite{Bhardwaj_Gujral_Wu_Zogheib_Maharaj_Becker_2024} and documentation practices in the NeurIPS Datasets and Benchmarks track~\cite{Yang_Liang_Zou_2023}, respectively. Both studies found that documentation comprehensiveness varied wildly, with~\citeauthor{Bhardwaj_Gujral_Wu_Zogheib_Maharaj_Becker_2024} noting specifically that although the most popular datasets generally include populated dataset cards, the vast majority of datasets overall do not. When documentation is provided, fields related to the data structure and function are significantly more detailed than those related to social, ethical, or contextual considerations (which are either minimally filled in or entirely absent). Despite this, \citeauthor{Holstein_Vaughan_III_Dudík_Wallach_2019} find that ``[industry practitioners] typically look to their training datasets, not their ML models, as the most important place to intervene to improve fairness in their products"~\cite{Holstein_Vaughan_III_Dudík_Wallach_2019}.

These results point to a critical need for empirical analysis designed to understand the motivations driving documentation tool development, and how those motivations do (or do not) lead to real-world adoption. Our is the first study that, to our knowledge, systematically reviews all available dataset documentation tools, assesses the implicit and explicit values they advance, and synthesizes findings into recommendations for the design and future integration of data documentation tools into widespread community practice.

\section{Methods}
\label{sec:methods}

To comprehensively study how researchers approached the design of dataset documentation tools, specifically in the fields of Computer Science (CS), Machine Learning (ML), HCI, and Ethics of Technology, we conducted a scoping review of written literature across several databases of interest. Our focus was on exploring the goals, motivations, concepts, and efforts towards integration behind these tools. We chose to use a scoping review for this study---including for identifying our research questions---given that, as opposed to a systematic review, we do not evaluate studies on the basis of quality nor use them to respond pre-registered hypotheses ~\cite{arksey_scoping_2005}. Instead, our research is exploratory: we aim to report on practices related to conceptualizing and designing dataset documentation tools; uncover themes across their use; and identify their labor and integration costs. To strengthen the robustness, transparency, and replicability of our procedures, we followed the PRISMA guidelines for scoping reviews~\cite{tricco_prisma_2018}. Lastly, we report our findings using reflexive thematic analysis~\cite{braun_thematic_2022}.

\subsection{Research Questions}
Following~\citeauthor{arksey_scoping_2005}, we began the study by stating our research questions. The first author then engaged in an initial review of a set of papers introducing dataset documentation tools. Subsequently, three authors read a small sample of key papers. Following this second round, we refined our initial set to generate the following three research questions:

\begin{itemize}
    \item \textbf{RQ1:} What were the goals and motivations behind the building of dataset documentation tools? (e.g., different tools may advance different values, different politics, potentially driving different goals)
    \item \textbf{RQ2:} How did creators of documentation tools conceptualize the dataset documentation process? (e.g., what are the different perspectives and approaches to documentation as reported by authors)
    \item \textbf{RQ3:} How do these tools connect and integrate with existing systems, regulation, or cultural norms?
\end{itemize}

\subsection{Data Collection}
The entire data collection process took place in four stages: database and repository searches, paper screening, eligibility checks, and data extraction and analysis. 

\subsubsection{Information Sources and Search Strategy}
In an effort to capture a wide net of resources in our review, and to specifically target key venues associated with the work of interest, we queried three established databases (ACM Digital Library, IEEE Xplore, and Science Direct), two domain specific repositories (ArXiv and ACL Anthology) and two major conferences (AAAI and NeurIPS). We included these conferences because of their standing in fields relevant for this study, and because they do not systematically deposit their contributions into any of the previously mentioned databases and repositories. 

We used a variety of search terms and term combinations to find tools (e.g. “dataset”, “documentation”), and nuanced those searches to match contexts of interest with descriptors like “transparency,” “accountability,” and “provenance.” These words emerged from our initial review of papers. We used these words in formal queries across abstracts, titles, author keywords, and sometimes over full texts depending on the affordances of each resource’s search engine. Additionally, following~\citeauthor{de_angelis_about_2021}, we explored grey literature using Google’s advanced search. To determine saturation, we used a five-page-noise rule, where we declared the search exhausted after five pages of results with no relevant hits. This generated 1,491 results across all sources databases, and 118 results across Google’s services, all collected in February 2025. This set was complemented with a round of snowballing, resulting in a total of 1,582 materials after removing duplicates. Our search only included English materials. The full queries for each resource are provided in Table \ref{sec:appendix-search-queries} in the appendix.

\subsubsection{Screening and Elegibility}
While scoping our research questions, we developed a set of inclusion/exclusion criteria. We included items that:

\begin{itemize}
    \item Proposed or advanced a dataset documentation tool 
    \item Proposed or advanced a documentation framework (to qualify, frameworks needed to include at least one dimension related to data documentation)
    \item Extended an existing dataset documentation tool
    \item Provided an empirical evaluation of a tool/framework in practice
\end{itemize}

Conversely, we excluded (1) items that only offered design recommendations and (2) case studies of the application of a dataset documentation tool or framework that did not include any evaluation or reflexive analysis.

One author used these criteria during a first “desk-reject” filtering, removing 1,408 items. We retained articles with ambiguous titles, abstracts, and keywords, leaving 115 items. Using random samples of 20 items, three authors calibrated decisions and reviewed papers for inclusion across three consecutive screening rounds with increasing Cohen Kappa agreement results ($\kappa$ = 0.29, $\kappa$ = 0.54, and $\kappa$ = 1.0, respectively). Following this last round, one of the authors screened the remaining items in the sample resulting in a final set of 59 items for data analysis. Figure \ref{fig_7_prisma_datasets} in the appendix outlines the process from search to final corpus.

\subsubsection{Data Extraction}
\label{sec:methods:data-extraction}
We retrieved quantitative information from our dataset to power several descriptive analyses. For each item that advanced a dataset documentation tool or framework, we noted (1) if the study was based on a needs assessment/user research study; (2) if a study was used, what type of study; (3) the stated audiences for the tool/framework; (4) if the paper mentioned a \textit{need} to involve stakeholders in tool development; (5) if stakeholders (e.g., users, policymakers) were \textit{actually} involved in the process; (6) the intended use of the tool; (7) if an evaluation of the proposals were included as part of implementation; (8) the degree of automation of each tool (manual, hybrid, fully automated)\footnote{\textit{Manual} tools were almost exclusively based on natural language provided by a human; \textit{hybrid} tools used natural language input from a human and to produce documentation artifacts; and \textit{automated} tools used machine readable information with almost no natural language input to produce documentation.}; (9) any integration with ecosystems of practice, other tools, or regulation; and (10) the terms used by authors to present their proposals (e.g., applications, datasheets, frameworks). Definitions for this categorization of different types of tools are included in Table \ref{fig:table-tools-definitions}. We also kept track of authors’ affiliations based on the information found in the paper (‘academia’, ‘industry’, ‘government’, ‘non-profit’, and ‘other’, which included independent researchers or members of the public), and used that classification to determine the distribution of creators across these sectors.

\begin{table*}[htbp]
\small
\renewcommand{\arraystretch}{1.5}
\centering
\begin{tabular}{p{2.0cm}p{8.0cm}p{3.0cm}}
\rowcolor{black}
\multicolumn{3}{c}{\color{white}\textbf{Tools Definitions}} \\ \hline
\multicolumn{1}{c}{\textbf{Type of tool}} &
\multicolumn{1}{c}{\textbf{Definition}} &
\multicolumn{1}{c}{\textbf{Items}} \\ \hline
    Application & Programs, web interfaces or digital tools provided to users as support to create dataset documentation & \cite{giner-miguelez_domain-specific_2023,schramowski_can_2022,giner-miguelez_datadoc_2023,petersenDataMaidYourAssistant2019,giner-miguelez_describeml_2022,halevy_goods_2016,sunMithraLabelFlexibleDataset2019,atli_tekgul_effectiveness_2022,novacek_ontology-supported_2024,romanOpenDatasheetsMachinereadable2024a,alencar_prov-dominoes_2024,arslan2019automatically} 
    \\ \hline
    Datasheet & Items named "datasheets", as well as proposals that provided structured batteries of questions that once answered became the documentation artifact & \cite{richards_methodology_2020,Microsoft_2022_Aether,srinivasan_artsheets_2021,papakyriakopoulos_augmented_2023,barman_datasheet_2023,marandi_datasheets_2025,gebru_datasheets_2021,heintz_datasheets_2023,Fact_Sheets_Arnold_Bellamy_Hind_Houde_Mehta_Mojsilović_Nair_Ramamurthy_Olteanu_Piorkowski_2019,rostamzadehHealthsheetDevelopmentTransparency2022a,adkins_method_2022,zheng_network_2022,mittal_responsible_2024,siddik_datasheets_2025}
    \\ \hline
    Framework & Structured, reflexive guidelines (sometimes in the form of questions) aimed at helping users define key information to include in a documentation artifact & \cite{fox_generative_2024,diaz_crowdworksheets_2022,Data_Cards_Pushkarna_Zaldivar_Kjartansson_2022,afzal_data_2021,bender_data_2018,luthra_data-envelopes_2024,horsman_dataset_2021,picardEnsuringDatasetQuality2020,bhardwajMachineLearningData2024a,alderman_tackling_2025,castelijns_abc_2020,The_CLeAR_Documentation_Framework_for_AI_Transparency_2024,holland_dataset_2018,chmielinskiDatasetNutritionLabel2022a,hutchinson_towards_2021}
    \\ \hline
    Markup format & Machine-readable format, usually implemented as an application in the form of a plugin or package & \cite{jainStandardizedMachinereadableDataset2024,osterweil_clear_2010,rondinaCompletenessDatasetsDocumentation2023,akhtarCroissantMetadataFormat2024,ahmad_toward_2024}
    \\ \hline
    Schema & Proposals that combined metadata information along with structured questions meant to become documentation, similar to datasheets & \cite{mcmillan-major_data_2024,mcmillan-major_reusable_2021}
    \\ \hline
    Study & Items describing an evaluation or study of a dataset documentation tool in practice & \cite{piorkowskiFieldStudyHumanCentered2024,boyd_datasheets_2021,crisanInteractiveModelCards2022a,Yang_Liang_Zou_2023,reid_right_2023,fabris_tackling_2022,Bhardwaj_Gujral_Wu_Zogheib_Maharaj_Becker_2024,heger_understanding_2022}
    \\ \hline
    Toolkit & Items providing reusable components such as templates or exercises that practitioners can tailor and use to plan and execute documentation over datasets & \cite{mcmillan-major_data_2023,mcmillan-majorLanguageDatasetDocumentation2023} \\ 
    \arrayrulecolor{black}\bottomrule
\end{tabular}
\vspace{1em}
\caption{Definitions and items for each type of dataset documentation tool identified in the systematic review. The table categorizes seven tool types: Application (13 items), Datasheet (14 items), Framework (15 items), Markup format (5 items), Schema (2 items), Study (8 items), and Toolkit (2 items). Classification was based on the language used by authors in each paper when introducing their proposals.}
\label{fig:table-tools-definitions}
\end{table*}

\subsection{Data Analysis}
We analyzed our corpus using by Braun and Clarke’s reflexive thematic analysis approach~\cite{braun_thematic_2022}. We chose this method since it let us recognize and reflect on the influence of our experiences as we navigated the data \cite{braun_one_2021}. Three authors from our team led the analysis, with each assigned a set of roughly 20 items from our final sample. This group of authors independently reviewed each article in their set and inductively developed and maintained codes definitions using the ATLAS.ti qualitative coding software. We maintained an open coding approach for the first round, developing 133 codes grouped into 26 clusters that we later mapped onto our research questions. We used this as a springboard to conceptualize, discuss, and refine themes iteratively across subsequent rounds. Themes were developed together with memos from each of us involved in the coding process, and with reference to representative quotes. We want to acknowledge that, in presenting these themes, we often directly quote authors from papers included in the sample and sometimes paraphrase them, often to benefit the flow of our arguments, and always with great attention not to add our own interpretation.

\subsection{Positionality of the Research}
One of the guiding questions for our review is if and how the Responsible AI community is advancing standards and norms around dataset documentation in light of its commitment to fairness, accountability, and transparency. While our community signals that these goals are important, the way individual researchers hold to them might differ and, as we mention in Section~\ref{sec:discussion}, there is no clear path for transforming these values into collective norms. We want to stress that neither our review nor our research intends this as a critique to the researchers or works we review, but rather as a reflection on how we can collectively move towards these goals. This reflection is a result of our experiences and intellectual commitments as authors. All four authors have formal training in HCI. The first author is an HCI and robotics researcher committed to critically studying sociotechnical systems by centering the experiences of groups greatly affected by automated technologies, yet with little influence or power over them. The second author explores the design of data-driven technologies that support historically marginalized groups. The third and fourth authors investigate the data work practices of practitioners developing LLMs. 
We also want to note that the vast majority of work included in our review is representative of a Western approach to dataset documentation. Therefore, it can only be representative of the way in which that epistemological position understands data production, consumption and dissemination, as well as its corresponding governing processes. As authors, we acknowledge that, while often holding critical views around how these processes and governance take place, we also partake in them. Consequently, this subjective tension is reflected in our analytic approach.

\section{Findings: Descriptive Statistics}
\label{sec:findings:descriptive-stats}

In this section, we provide descriptive statistics of our corpus regarding the types of documentation tools we found, how tools’ prevalence and level of automation has progressed over time, and the audiences they consider. 

\subsection{Diversity and Evolution}
\begin{figure}[ht]
    \centering
    \includegraphics[alt={Figure 2 contains six stacked vertical bar graphs arranged in two rows, showing the count of documentation tools by degree of automation (Manual, Hybrid, Automated) across years 2010 to 2025. Each panel represents a different tool type: toolkit, framework, application, datasheet, markup format, and schema. The vertical axis for each graph shows Count, ranging from 0 to 5. The horizontal axis shows Year. The data are summarized as follows. Toolkit panel: 2023 has 2 Hybrid tools, all other years have 0. Framework panel: 2018 has 1 Manual and 1 Hybrid; 2020 has 1 Manual and 1 Automated; 2021 has 2 Manual and 1 Hybrid; 2022 has 2 Manual and 1 Hybrid; 2023 has 1 Manual, 2 Hybrid, and 1 Automated; 2024 has 1 Manual and 1 Automated; all other years have 0. Application panel: 2016 has 1 Hybrid; 2019 has 1 Manual and 2 Automated; 2022 has 2 Hybrid and 1 Manual; 2023 has 2 Automated; 2024 has 1 Manual and 3 Hybrid; all other years have 0. Datasheet panel: 2019 has 1 Manual; 2020 has 1 Hybrid; 2021 has 2 Manual; 2022 has 3 Manual and 1 Hybrid; 2023 has 3 Manual; 2024 has 1 Hybrid; 2025 has 1 Hybrid; all other years have 0. Markup format panel: 2010 has 1 Hybrid; 2023 has 1 Manual; 2024 has 2 Hybrid and 1 Automated; all other years have 0. Schema panel: 2021 has 1 Manual; 2024 has 1 Manual; all other years have 0.}, width=\linewidth]{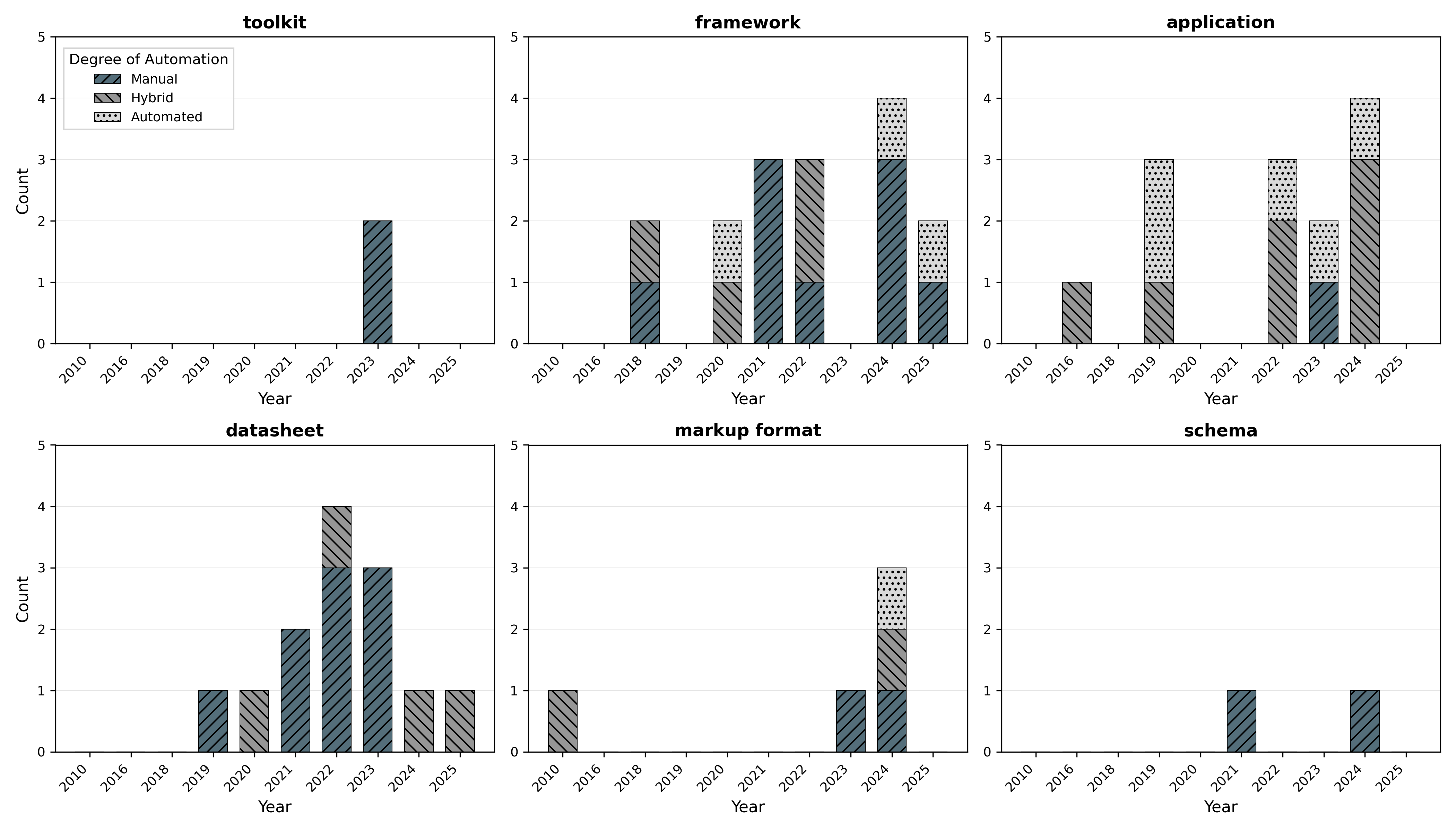}
    \caption{Distribution over time of different approaches to the production of dataset documentation across the six categories of tools in our sample}
    \label{fig_2_type_of_tool_automation_over_time}
\end{figure}

We found that the total number of dataset documentation tools has increased over time, with a noticeable concentration of new tools between 2022 and 2024 (31 out of 51 proposals). The types of tools and terms used describe them varied; we ultimately identified seven: toolkits, frameworks, applications, datasheets, markup formats, and schemas. Definitions for each tool can be found in Table~\ref{fig:table-tools-definitions}. \textit{Toolkits}, \textit{datasheets}~\cite{gebru_datasheets_2021,heintz_datasheets_2023}, \textit{frameworks}~\cite{afzal_data_2021,The_CLeAR_Documentation_Framework_for_AI_Transparency_2024}, and \textit{schemas}~\cite{mcmillan-major_data_2023} offered structured guidance to help users document. \textit{Datasheets}, for example, entailed batteries of questions, while \textit{schemas} combined metadata with questions. \textit{Applications} and machine-readable \textit{markup formats} facilitate the automation of the documentation process. Frameworks (n=16), applications (n=13) and datasheets (n=13) are the most popular types, amounting to 82\% of our total sample. Figure \ref{fig_2_type_of_tool_automation_over_time} shows these distributions. 

\begin{figure}[ht]
    \centering
    \includegraphics[alt={Figure 1 is a stacked vertical bar graph showing the count of documentation tools by type from 2010 to 2025. The vertical axis shows Count, ranging from 0 to 14. The horizontal axis shows Year. The total number of tools increases significantly over time, from 1 tool in 2010 to a peak of 13 tools in 2024, before declining to 3 in 2025. The data are summarized in the following table: Count of Documentation Tools by Type and Year. Year 2010: Application 0, Datasheet 0, Framework 0, Markup Format 1, Schema 0, Toolkit 0, Total 1. Year 2016: Application 1, Datasheet 0, Framework 0, Markup Format 0, Schema 0, Toolkit 0, Total 1. Year 2018: Application 0, Datasheet 0, Framework 2, Markup Format 0, Schema 0, Toolkit 0, Total 2. Year 2019: Application 3, Datasheet 1, Framework 0, Markup Format 0, Schema 0, Toolkit 0, Total 4. Year 2020: Application 0, Datasheet 0, Framework 3, Markup Format 0, Schema 0, Toolkit 0, Total 3. Year 2021: Application 0, Datasheet 2, Framework 3, Markup Format 0, Schema 1, Toolkit 0, Total 6. Year 2022: Application 3, Datasheet 4, Framework 2, Markup Format 1, Schema 0, Toolkit 0, Total 10. Year 2023: Application 2, Datasheet 3, Framework 0, Markup Format 2, Schema 0, Toolkit 1, Total 8. Year 2024: Application 4, Datasheet 1, Framework 4, Markup Format 3, Schema 1, Toolkit 0, Total 13. Year 2025: Application 0, Datasheet 1, Framework 2, Markup Format 0, Schema 0, Toolkit 0, Total 3.}, width=\linewidth]{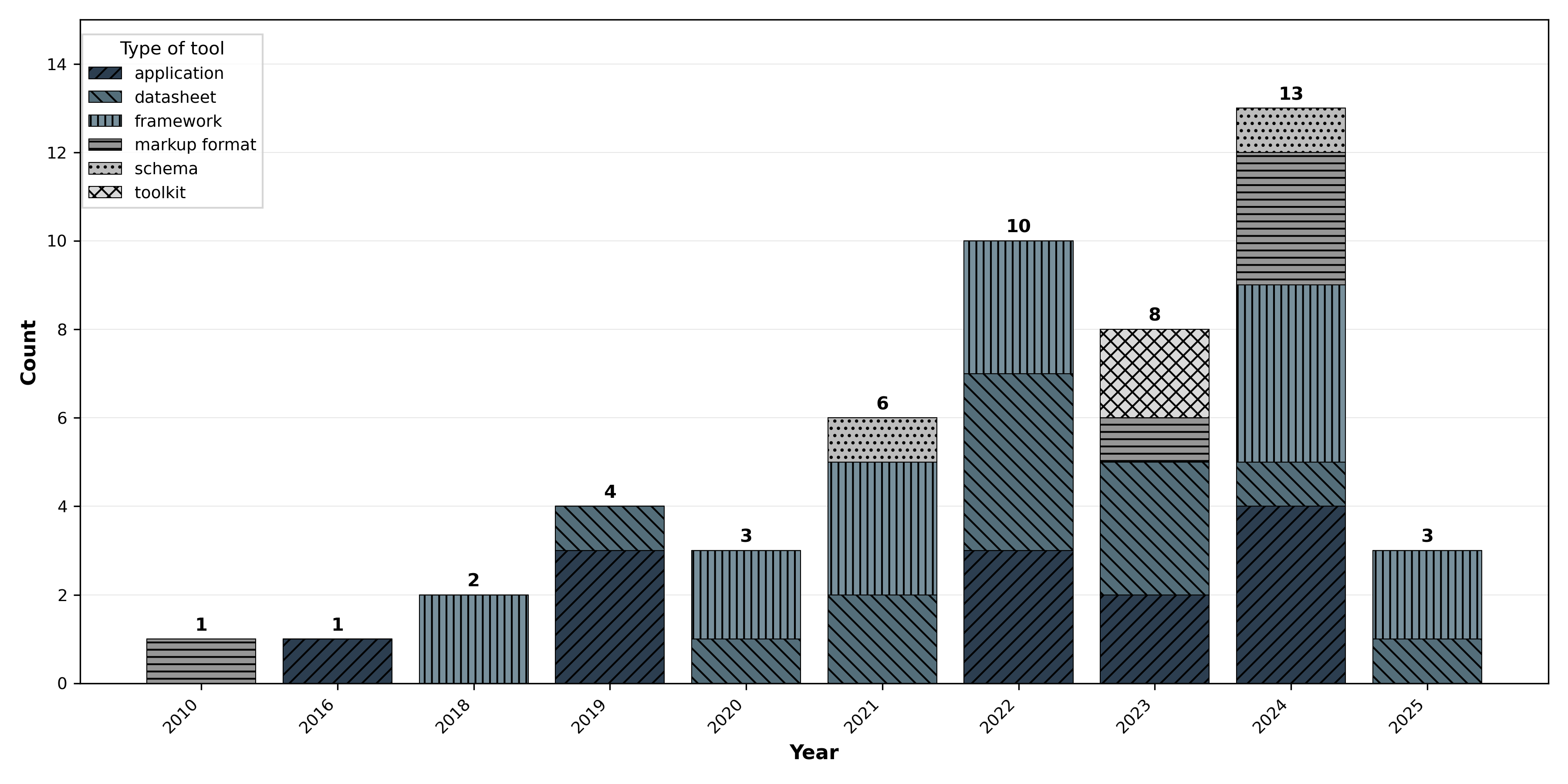}
    \caption{Distribution of different types of tools over time based on the sample in our corpus. It is worth noting that 2025 is an outlier year given our sample was finalized in March of that year.\protect\footnotemark}.
    \label{fig_1_different_tools_over_time}
\end{figure}

We also note a prevalence of tools that require manual input to produce documentation (n=25). However, recent years show an increase in hybrid tools that use natural language to automatically produce documentation (n=17) and automated tools that use machine readable information to generate documentation (n=9), a trend that matches advances in large language models. Figure \ref{fig_1_different_tools_over_time} visualizes these trends.

The increase in both volume and types of tools over the past decade or so appears to signal a high diversity in creators' conceptualizations of documentation structure and role within larger processes. Meanwhile, an increase in automated tools may simply coincide with the advent of mainstream, more capable LLMs.

\footnotetext{Distribution of different types of dataset documentation tools over time. Data shows a noticeable increase in the production of these tools between 2021 and 2024}

\subsection{Creators and considered Audiences}
To explore our first research question (goals and motivations of tool development), we first looked at author’s affiliations. We found that half of the papers in our sample were authored solely by researchers in academia (31 papers, 52.54\%), and a majority of papers included at least one author from academia (46 papers, 77.97\%) . Researchers in industry contributed to 27 papers (45.76\%), and participated in 12 papers (20.34\%) with members from other groups. Participation in papers in our sample from other groups included government (7 papers, 11.86\%), non-profits (6 papers, 10.17\%), and other (1 paper, 1.69\%). We found a similar distribution when looking at authors per affiliation across all papers (with total authors as the denominator). In order: academia (180 authors, 51\%), industry (139 authors, 39.38\%), government (20 authors, 5.67\%), non-profits (13 authors, 3.68\%), and others (1 author, 0.28\%). 
Our analysis also highlighted that documentation tools’ creators used diverse terms (e.g., dataset creators, dataset experts, and data curators) for defining their target user profile. Dataset creators emerged as the term that creators most commonly used to identify their audience. While some tool types (e.g., datasheets) tended to identify a wide variety of audiences, most failed to engage users in the design process. Visualizations for this and other distributions can be found in appendix \ref{sec:appendix-descriptive-analyses}.

\begin{figure*}[ht]
    \centering
    \includegraphics[alt={Figure 3 is a Sankey diagram showing the progression of stakeholder engagement across three stages for 51 documentation tools. The diagram flows from left to right, with bands representing the flow of tools between stages. Starting with All Tools (n=51), the diagram shows: Stage 1 Mention divides into Stakeholders Not Mentioned (n=30, 58.8 percent) and Stakeholders Mentioned (n=21, 41.2 percent). Stage 2 Involvement shows that of the 21 tools that mentioned stakeholders, 11 (21.6 percent of total) had Stakeholders Not Involved and 10 (19.6 percent of total) had Stakeholders Involved. Stage 3 Integration shows that of the 10 tools with stakeholders involved, 8 (15.7 percent of total) had Stakeholder input not Integrated and 2 (3.9 percent of total) had Stakeholder input Integrated. This demonstrates a progressive narrowing from mention to involvement to integration of stakeholder input.}, width=\linewidth]{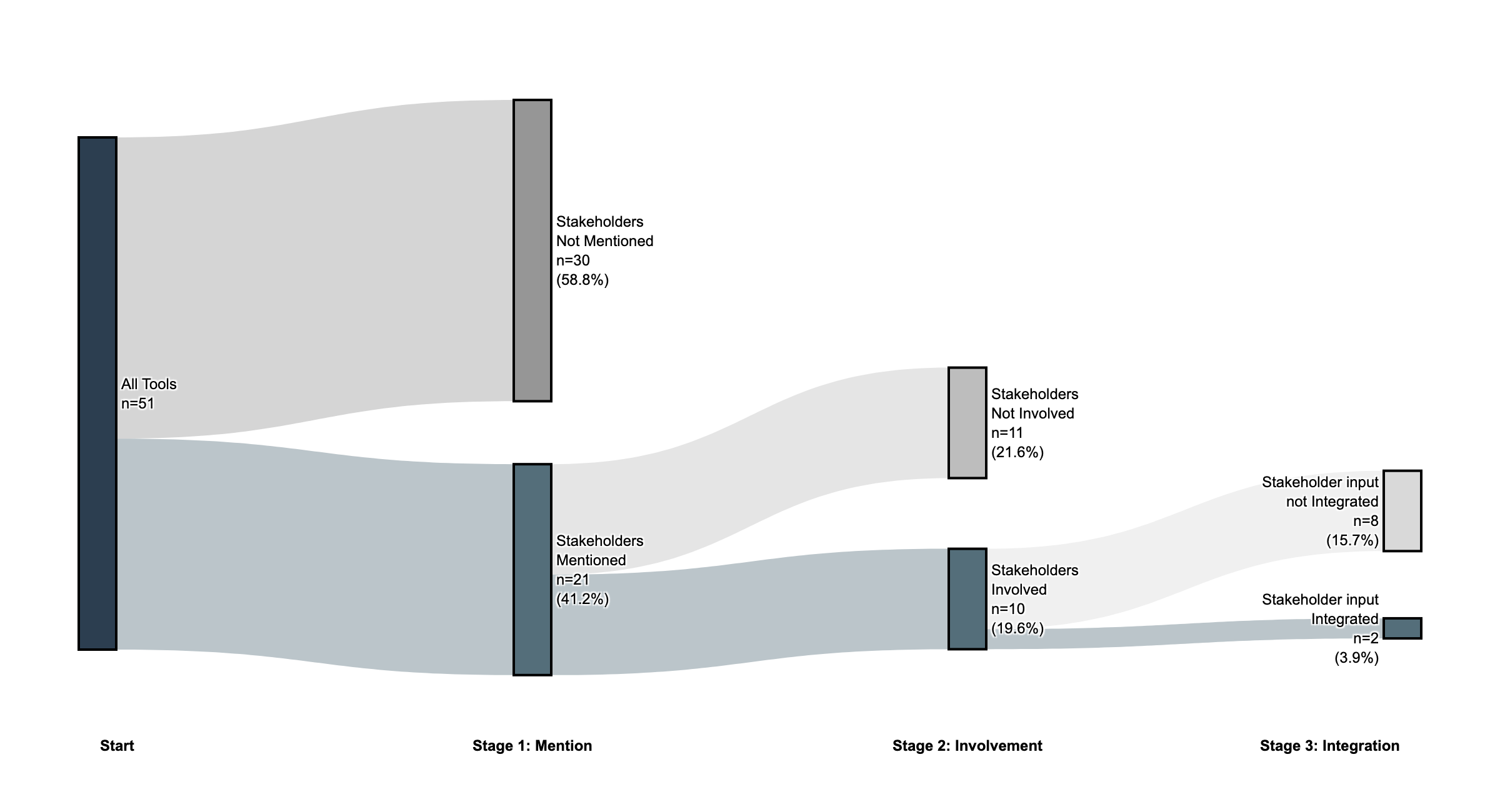}
    \caption{Comparative analysis of stakeholder engagement and integration pipelines including the percentage of proposals that specifically mentioned stakeholders as part of the design/use stages compared to the number of proposals from within that group that included any stakeholder in the process. It also showcases the percentage of proposals that included stakeholders during the design/testing stages compared to the number of proposals that included concrete features or guidance towards integration to stakeholders.}
    \label{fig_3_stakeholder_integration_sankey}
\end{figure*}

We further explored this gap between designers and stakeholders, and found that more than half of the tools in our dataset (n=30) did not mention engaging relevant stakeholders during the design process. Of the remaining tools, most only described eliciting views without explicitly integrating them into tool design. Only two items in our sample described eliciting \textbf{and} integrating stakeholder input (See Figure \ref{fig_3_stakeholder_integration_sankey}).  

\section{Findings: Thematic Analysis}
\label{sec:findings:thematic-analysis}

The qualitative analysis of our corpus foregrounded four critical aspects of how tools’ creators conceptualize dataset documentation tools: a disagreement on how to operationalize the value of documentation; a tendency to design decontextualized tools disconnected from user experiences; an increase of undiscussed labor demands, even when moving towards automation; and a view of tools integration as an aspirational future endeavor. Each of these aspects illustrates how creators assumptions about productive documentation practices impact the adoption of dataset documentation tools.

\subsection{From Reflection to Repair: A Spectrum of Views on the Value of Documentation}

Our analysis suggests that a prevalent goal motivating the design of documentation tools is to counteract the \textit{“undesired consequences and negative downstream effects in the whole machine learning pipeline due to data issues”}~\cite{siddik_datasheets_2025}. Indeed, most creators agreed that \textit{“the quality of a dataset used to build a model will directly influence the outcomes it produces”}~\cite{holland_dataset_2018} and positioned documentation as the mechanism for ensuring the transparency and accountability of dataset work. In particular, tools’ creators argued that only adequately documented data \textit{“can be appropriately used”}~\cite{fabris_tackling_2022} given that high-quality documentation \textit{“is a lever that enables accountability”}~\cite{The_CLeAR_Documentation_Framework_for_AI_Transparency_2024} and \textit{“directly impacts the transparency, reliability, and reproducibility in the field of data-driven research”}. However, authors differed on how, in practice, documentation might achieve these benefits. Views included: (1) that documentation serves as personal reflection for dataset creators, (2) that documentation would allow downstream users to scrutinize datasets, and (3) that documentation would repair broken dataset infrastructure. Given that most tools were designed by two groups---researchers in academia and in industry---one might expect significant overlap around the value of documentation. However, our findings show this assumption does not hold. These perspectives profoundly shaped tools’ documentation attributes, target audience, and intended impact. Moreover, they foregrounded the need for evidence supporting the assumption that documentation tools can increase datasets’ and AI’s transparency and accountability.

\subsubsection{Eliciting Dataset Creators' Careful, Personal Reflection}

Some proposals described documentation as the means to facilitate personal reflection by dataset creators about dataset-related decisions. As ~\citeauthor{gebru_datasheets_2021}, the creators of 'Datasheets for Datasets' explained, the argument behind this view is that personal, careful reflection on dataset management increases \textit{“transparency and accountability within the machine learning community, mitigate unwanted societal biases in machine learning models, facilitate greater reproducibility of machine learning results, and help researchers and practitioners to select more appropriate datasets for their chosen tasks”}~\cite{gebru_datasheets_2021}. Tools advancing this view relied heavily on reflection-eliciting guidance (e.g., questions) particular to the tool's domain or end-goal. For example, in the context of documenting annotators' histories and experiences, Crowdworksheets~\cite{diaz_crowdworksheets_2022} included considerations for documenting \textit{“how much annotator’s identity, lived experience, and prior knowledge of a problem space matters for the task at hand, and how it impacts what the resulting dataset is intended to capture.”}

Tools holding this view often struggled to balance between reflection and pragmatism. For example, ‘Datasheets for Datasets’, which is not domain-specific, prioritizes reflection to the point of advising against automation (as this could \textit{“run counter to our objective of encouraging dataset creators to carefully reflect on the process of creating, distributing, and maintaining a dataset”}~\cite{gebru_datasheets_2021}). Recognizing the overhead that reflection can create for users, other tools (e.g,~\citeauthor{marandi_datasheets_2025} and~\citeauthor{luthra_data-envelopes_2024}) offered a mix of practical, time-efficient sections (e.g., concrete checklists and machine-readable formats) and reflection-eliciting sections (e.g., the dataset creators’ positionality~\cite{luthra_data-envelopes_2024}).

The reflection viewpoint also saw documentation as an activity that needed to take place \textit{“prior to any data collection”}~\cite{gebru_datasheets_2021} so as to \textit{“promote a culture of responsible data stewardship from the very beginning of the research lifecycle”}~\cite{luthra_data-envelopes_2024}. As such, tools had components for dataset creators to conduct “a thorough structural analysis of the dataset during the planning phase”~\cite{luthra_data-envelopes_2024}, thereby identifying data ethics issues early on.

Despite the emphasis on reflection-oriented components and early documentation, we found that, in their initial proposals, tools under this view did not provide evidence of their impact. Specifically, they tended to miss evaluations that their reflection-based design choices increased users' reflexivity capacity and, more importantly, improve the transparency, reliability, accountability, and ethical integrity of datasets. In some cases, however, other authors offered evidence for tools with a higher level of adoption (e.g., \cite{Data_Cards_Pushkarna_Zaldivar_Kjartansson_2022, boyd_datasheets_2021})

\subsubsection{Offering Dataset Consumers the Ability to Scrutinize Datasets}

The next operationalization frame we identified sought to provide dataset consumers with scalable means for intensely scrutinizing datasets. As the creators of the ‘Dataset Nutrition Label’ explained, this view relies on the idea that, \textit{“to improve the accuracy and fairness of AI systems, it is imperative that data specialists are able to more quickly assess the viability and fitness of datasets, and more easily find and use better quality data to train their models”}~\cite{holland_dataset_2018}. As such, tools under this perspective included functionality for augmenting the visibility of critical dataset attributes, like those related to fairness or bias concerns~\cite{giner-miguelez_domain-specific_2023}, and details of \textit{“operations conducted during the data preparation, cleaning, and quality analysis phases in a typical AI life cycle”}\cite{afzal_data_2021}. While these tools also included qualitative attributes, their emphasis was on factual rather than reflective information such as metadata, provenance descriptions~\cite{holland_dataset_2018}, quality and remediation assessments,~\cite{afzal_data_2021}, and potential bias issues~\cite{giner-miguelez_datadoc_2023}. In a few cases, tools included a comments section where consumers and creators could share concerns~\cite{holland_dataset_2018}. \citeauthor{holland_dataset_2018} explain that quantitative combined with factual qualitative information can better motivate reflection on the consumer side: \textit{“[the label] enables the data specialist to better understand and ascertain the fitness of a dataset by scanning missing values, summary statistics of the data, correlations or proxies, and other important factors. As a result, the data specialist may discard a problematic dataset or work to improve its viability prior to utilizing it.”}~\cite{holland_dataset_2018}. 

Corpus tools under this view often missed identifying reasons for requiring specific quantitative or qualitative documentation attributes. Further, similarly to the reflective tools discussed above, tools emphasizing consumer scrutiny did not provide evidence that their design choices led to easier, quicker, or more effective dataset analysis. Thus, there is a pressing need to demonstrate how facilitating dataset scrutiny may eventually lead to increased transparency and accuracy.

\subsubsection{Facilitating the Repair of Broken Dataset Infrastructure} 

A third set of tools, championed documentation as the means to repair incomplete and messy data ecosystems. Work from ~\citeauthor{hutchinson_towards_2021}, for example, argued that data ecosystems are increasingly breaking down, creating datasets that are \textit{“poorly maintained, lacking in answerability, and have opaque creation processes”}. As such, they proposed to use documentation as \textit{“a deliberative and intentional methodology, rather than the post-hoc justifications that are sometimes observed when datasets are developed hastily or opportunistically.”} In contrast with this view of repair as intentional and transversal actions, most other tools under the repair view proposed repair as quick, seamless fixes taking place after dataset creation. For example, to solve the problem of datasets in an infrastructure residing in different storage systems and taking \textit{“a variety of forms, such as structured files, databases, spreadsheets, or even services that provide access to the data”}, ~\citeauthor{halevy_goods_2016} proposed harnessing datasets’ metadata and users’ comments to organize datasets at scale. The tool's goal was to work \textit{“in the background, in a non-intrusive manner”}. Similarly, the authors of~\cite{fox_generative_2024} addressed the complexity of identifying multiple versions of the same dataset in a data repository by proposing a Generative AI-powered framework that produced documented benchmarks for data versioning without users’ intervention. Other quick fixes entailed creating artifacts (e.g.,  rubrics~\cite{Bhardwaj_Gujral_Wu_Zogheib_Maharaj_Becker_2024} and lightweight documentation formats~\cite{fabris_tackling_2022}) that could help quickly compensate for poor-quality documentation. As we noted in section \ref{sec:findings:descriptive-stats}, the rise in these types of tools coincides with breakthroughs in LLM technology. We discuss the potential ramifications of this move towards automation in the context of this "quick-fix" approach.

Design-wise, tools pursuing repair as quick fixes tended to select documentation attributes that were easy to record and machine-readable (e.g., metadata~\cite{halevy_goods_2016}, lineage and transformations~\cite{fox_generative_2024}, and semantic descriptions~\cite{ahmad_toward_2024}). However, tools often failed to discuss the implications of focusing on some documentation elements at the expense of others. For example, ~\citeauthor{fabris_tackling_2022} missed discussing the impact of compacting the ‘Datasheets for Datasets’ documentation format, which left behind the reflection-based questions that characterize that Datasheets original design. Furthermore, similarly to tools in the other two perspectives, we found tools did not provide evidence of how different repair strategies managed to repair dataset ecosystems and increased accountability and transparency. 

\subsection{The Role of De-contextualization in Devaluing the Dataset Infrastructure}

As~\citeauthor{hutchinson_towards_2021} argued, data infrastructure is continuously undergoing a \textit{“systemic devaluation of dataset work”} and of data work more broadly~\cite{hutchinson_towards_2021}. Our analysis highlights that avoiding such devaluing was critical for many corpus authors such as ~\cite{Microsoft_2022_Aether}. As they explained, their tool strives to increase the visibility of existing data work for all possible actors involved: \textit{“someone completely unfamiliar with the dataset would be able to make an informed decision about whether and how to use this dataset responsibly”}~\cite{Microsoft_2022_Aether}. However, our quantitative findings indicated that, as most tools broadened the scope of who can document, they did so without including users and stakeholders in the design process (n=30, 58.8\%). Our thematic analysis stressed how this disconnection from users ran the risk of further complicating and devaluing dataset production.

\subsubsection{Wide-encompassing tools, narrow-encompassing contexts}

In studying how authors reported on the creation of their proposals, we observed a tendency to prioritize domain-agnostic characteristics. This trend matches our quantitative finding that the most open-ended, least constrained tools---datasheets and frameworks---comprised a majority of the total tools in our sample (29 out of 51). For example, ‘Data Cards’ framed their approach to documentation as \textit{“an underlying framework for transparency reporting for domain and fluency-agnostic readability and scaling in production contexts.”}~\cite{Data_Cards_Pushkarna_Zaldivar_Kjartansson_2022} The domain-agnostic characteristic, we found, was also relevant for creators leveraging these particular tools: \textit{“This paper focuses on Datasheets: a technique- and domain-agnostic, lay-language context document for training data. Datasheets are versatile: they can be taught early in ML education to students who will go on to work in diverse domains using a variety of techniques”}~\cite{boyd_datasheets_2021}.

Often, these context-independent design approaches attended to a wide range of goals, overlaid across several stages of the dataset production process, and cutting across domains of application. However, we found that, even when creators recognized that \textit{“…one size does not fit all”}, they often passed the burden of contextualization to their users: \textit{“This work has demonstrated that although FactSheets will contain some common elements, different FactSheets will generally contain different information, at different levels of specificity, depending on domain and model type. They will also contain different information for different industries and the different regulatory schemes within which these industries operate”}~\cite{Fact_Sheets_Arnold_Bellamy_Hind_Houde_Mehta_Mojsilović_Nair_Ramamurthy_Olteanu_Piorkowski_2019}. Similarly, the surrounding context was at times obscured by the pursuit of generality with few mentions of specific institutional arrangements, data governance structures, and relevant regional/national regulation of relevance to future users.

Our analysis highlights that more contextualized tools often built upon less contextualized ones. For example, ‘Artsheets for Datasets’ focuses \textit{“…specifically on the unique considerations (such as social, legal, cultural, historical, and environmental factors) that arise in the development of art datasets.”} Along with providing potential users with relevant contextual information to the field of art data, authors provided more detailed implementation guidance to counter the \textit{“common presumption of “one-size-fits-all” ethics checklists in the field by reinforcing the principle that ethical frameworks must be carefully adapted to each use…”}~\cite{srinivasan_artsheets_2021}. We found that tools that narrowed the domain and context of application thus presented a clearer picture of use along with responding to specific needs (e.g., terminology, standards, etc.), signaling the usefulness of their approach. However, creators of these domain-specific tools also stressed that a narrower focus came at the cost of reducing the utility of their tools for other domains (e.g., \textit{“focus on healthcare datasets may restrict its broader applicability to other domains”}~\cite{siddik_datasheets_2025}), and limited the possibility of comparison across datasets (e.g., \textit{“Future work requires a more principled approach for extending and adapting Data Card templates without compromising comparability”} ~\cite{Data_Cards_Pushkarna_Zaldivar_Kjartansson_2022}). This highlights an unresolved tension around the right scope for dataset documentation tools.

\subsubsection{Scopes of stakeholder engagement}

As we mentioned before, few tools engaged stakeholders in their design and integration process. Our thematic analysis, however, stresses that almost half of our corpus tools did discuss the importance of considering stakeholders (e.g., users, consumers and regulatory institutions) at some point of the tools’ creation timeline. Further, our analysis highlighted that the high variation of terms to describe stakeholders (e.g., “Data scientists”~\cite{Fact_Sheets_Arnold_Bellamy_Hind_Houde_Mehta_Mojsilović_Nair_Ramamurthy_Olteanu_Piorkowski_2019,chmielinskiDatasetNutritionLabel2022a,crisanInteractiveModelCards2022a,holland_dataset_2018,mcmillan-majorLanguageDatasetDocumentation2023,romanOpenDatasheetsMachinereadable2024a}, “data practitioners”~\cite{afzal_data_2021,sunMithraLabelFlexibleDataset2019,picardEnsuringDatasetQuality2020}, “data creators”~\cite{arslan2019automatically,bhardwajMachineLearningData2024a}, and “dataset creators”~\cite{Yang_Liang_Zou_2023,srinivasan_artsheets_2021,schramowski_can_2022,rostamzadehHealthsheetDevelopmentTransparency2022a,rondinaCompletenessDatasetsDocumentation2023}) was accompanied by a lack of definition of such terms and specifications of the date life-cycles where these roles operate. This inconsistency highlights a lack of collective agreement around foundational categories and roles in the dataset production chain, which hinders the identification and engagement of critical stakeholders.

Creators’ tendency to draw inspiration from other fields, we found, can also exacerbate issues around stakeholder engagement. For example, Factsheets were \textit{modeled after a supplier’s declaration of conformity (SDoC) [...] a document to “show that a product, process or service conforms to a standard or technical regulation, [...] used in many different industries and sectors including telecommunications and transportation} \citeauthor{Fact_Sheets_Arnold_Bellamy_Hind_Houde_Mehta_Mojsilović_Nair_Ramamurthy_Olteanu_Piorkowski_2019}. Authors, however, do not mention how users of SDoC's may have contributed to translating their use to the realm of datasets nor what limitations this ‘modeling-after’ might have. Interdisciplinary connections can produce interesting ideas while broadening---and possibly obscuring---the universe of relevant parties to include in the process. Without close understanding of how documentation tools operate in other fields and what stakeholders make that use successful, interdisciplinary connections can prove challenging. We return to the tradeoffs of interdisciplinary work in the next section.

\subsection{The Taxing Yet Undiscussed Labor Demands of Documentation}
\label{sec:findings:thematic-analysis:labor}

Our analysis stressed how, whether manual, hybrid, or fully automated, documentation tools impose labor demands upon practitioners in the form of time, access to institutionally-bound information, and interpretation. Further, we found that tools tended to place the responsibility of documenting on individuals rather than institutions. Yet, the authors in our corpus rarely discussed the labor demands of documentation and their broader implications. Finally, we observed an increasing shift to automated documentation tools (e.g., automating data extraction, incorporating LLMs into the data documentation pipeline, and hybrid approaches) without recognizing or discussing how it could reshape the nature and distribution of documentation labor. 

\subsubsection{Information Access and Dataset Interpretation: Undiscussed Forms of Labor}

Several authors in our corpus explicitly recognized that documenting can be time-demanding (e.g., it can take between six to twenty-four hours to create a FactSheets template~\cite{piorkowskiFieldStudyHumanCentered2024}), require diverse forms of human engagement (e.g., “active engagement with regulators and corporations”~\cite{jainStandardizedMachinereadableDataset2024}, and “work with experts in other domains such as anthropology, sociology, and science and technology studies” ~\cite{gebru_datasheets_2021}), and needs institutional incentives~\cite{diaz_crowdworksheets_2022}. However, it was less common for them to discuss these and other demands in terms of labor impositions for users. Our analysis highlighted the intensity and implications of two potentially taxing yet undiscussed labor demands of documentation tools: access to institutionally-regulated information, and interpretation of dataset work.

We found that many of the documentation tools in our corpus assumed that users would have seamless access to institutionally-regulated information. For example, ‘Datasheets for Datasets’ asks \textit{“Who was involved in the data collection process (for example, students, crowdworkers, contractors) and how were they compensated (for example, how much were crowdworkers paid)?”}~\cite{gebru_datasheets_2021}, which assumes users can easily access institutional policies or human resources records. Similarly, in the ‘Data Labeling’ use case, the Croissant-RAI vocabulary requires detailed information about annotation processes, including annotator demographics, instructions, and devices used. As the authors emphasized, documenting these attributes is essential for assessing label quality, but it also presupposes that the institutions where users operate have in place infrastructures that record and store this information in a systematic way~\cite{akhtarCroissantMetadataFormat2024}.

Our analysis also stressed that many documentation tools demand ethical interpretive labor. While some interpretation is a routine aspect of knowledge work, the ethical interpretive labor required by documentation tools constitutes a substantially more demanding activity. It requires practitioners---who often come from a technical background and lack ethics training---to make subjective but normative, anticipatory, and context-specific judgments about potential dataset implications. For example, tools such as Crowdworksheets asked users to judge the implications of datasets involving cultural or social contexts they might not fully understand. The tool includes questions such as, \textit{“does the dataset contain data that, if viewed directly, might be offensive, insulting, threatening, or might otherwise cause anxiety? If so, please describe why”}~\cite{diaz_crowdworksheets_2022}, and \textit{“Are there certain perspectives that would be harmful to include? If so, how did you screen these perspectives out?”}~\cite{diaz_crowdworksheets_2022}. As Heger et al. found, such interpretation could result in invisible and unevenly distributed labor burdens to users: \textit{“practitioners creating dataset documentation had trouble making connections between the questions they were asked to answer and their RAI implications and difficulty providing information that someone unfamiliar with their datasets would need to understand the data”}~\cite{romanOpenDatasheetsMachinereadable2024a}.Recognizing this form of interpretive work is therefore essential for adapting documentation tools so they provide appropriate guidance, distribute responsibilities across institutions rather than individuals, and avoid imposing additional unplanned workloads.

These examples illustrate that documentation requires more than the visible time of writing and coordinating. It also demands the less visible---and thus, less acknowledged and compensated---labor of figuring out how to navigate institutional regulations to access information requested by tools along with interpreting, translating, and judging datasets. These two forms of labor also shed light on the tendency for tools to center individual users as solely responsible for successful documentation, ignoring that the labor of documentation goes beyond the individual. Institutional buy-in and cooperation with other key stakeholders in adopting and enforcing the use of documentation tools is critical to ensure its adoption. If records are scattered or poorly maintained, or if there is a lack of expert support on domain-specific and ethical issues, documentation tools can generate additional workloads, thereby reinforcing users’ perceptions of documentation as burdensome. 

\subsubsection{The Move Towards Automation: Risks Of Disguising Labor}
Relatedly, several articles in our corpus proposed hybrid or fully automated approaches to documentation for three purposes: scalability through saving time and resources, compliance with institutional policies, and expanding the scope of documentation analysis. However, we found that the move towards automation could transform and further obscure labor demands to users, engendering risks that are harder and more resource-consuming for users. 

Our analysis highlighted how enabling scalable solutions that can cater to an ever-growing scale of current AI technologies was a priority for some tools moving towards automation. As ~\citeauthor{halevy_goods_2016} explained, automation can achieve this goal by preventing effort duplication and reducing manual work during documentation analysis: \textit{“if we can identify natural clusters to which datasets belong, then not only we can provide the users with a useful logical-level abstraction that groups together these different versions but we can also save on metadata extraction.”} As the creators of ‘DataDoc Analyzer’ further argued, automation can reduce documentation processing to fifty and sixty seconds for unseen documents and to up to twenty to twenty-five seconds~\cite{giner-miguelez_datadoc_2023}. Automation, other authors argued, could also relieve users from the need to manually record objective, traceable facts: \textit{“Code repositories adapted for use in AI development can record key facts about training data and model versioning”}~\cite{piorkowskiFieldStudyHumanCentered2024}. Our analysis showed that, often the time and resource demands of automation did not fully disappear but shifted to other forms or groups. In the case of the ‘Data Nutrition Label’, for example, creators discussed how their automated analytics module would require community work to prevent failure: \textit{“The veracity and usefulness of the Ground Truth Comparison module depends on the accuracy of the “Ground Truth” dataset [..] a mitigating step is to build Labels for ground truth datasets themselves. If these Labels include community feedback and comment modules, dataset authors can address the issues directly”}~\cite{holland_dataset_2018}.

Another group of creators in our corpus framed automation in documentation as a means of ensuring a systematic and sustainable compliance with emerging regulatory and organizational policies. In this sense, automation is positioned both as cost-saving and efficient, as well as a mechanism for embedding policy requirements into documentation workflows. For instance, the authors of ‘DataDoc Analyzer’ pointed out that \textit{“recent public regulatory initiatives such as the European AI Act and the AI Right of Bills [..] call for this documentation to be easy to understand by non-experts to bridge the gap between technology and end users”}~\cite{giner-miguelez_datadoc_2023}. Similarly, the creators of ‘Open Datasheets’ emphasized that automation can \textit{“ensure that open datasets align with their Responsible AI policies,”} streamlining much of the evaluation process. Our analysis stressed that such abidance by regulation through automation can also shift forms of labor to other actors. As~\cite{romanOpenDatasheetsMachinereadable2024a} explain,  \textit{“human review will still be necessary to make decisions based on organizational policies.”}. Further, relying on automation to abide by regulations could also force users to navigate obscured decisions about datasets and their attributes. For example, the creators of ‘Goods’ explain how automating dataset organization based on metadata and users’ comments depends on an automatic exclusion of \textit{“many types of uninteresting datasets”}, and a normalization of obvious redundancies. These automatic and hidden decisions could generate further unanticipated and hard-to-manage forms of labor for users. 

Other authors positioned automation, often through LLMs, as the pathway to extend documentation and dataset analytic possibilities beyond human capacity. Fox et al., for example, highlighted how documenting multi-hop dataset transformations is limited by human capacity and positioned the use of LLMs as a critical tool to navigate this barrier: \textit{“Recently, Large Language Models (LLMs) have been showing promising results in annotation and semantic tasks such as column type detection, entity matching, benchmarking table union search, and more”}~\cite{fox_generative_2024}. Likewise, Holland et al. suggested introducing more intelligence to next ‘Dataset Nutrition Label’ iterations to better detect subtle biases in datasets that humans might not recognize: \textit{“Analyses of machine bias indicate that zip codes often proxy for race, but many others proxies still exist, especially as the models themselves approach levels of complexity that are difficult or impossible for humans to comprehend and new or unexpected proxies emerge. Integrating new methods or tools to help identify proxies will be important to the industry ...”}~\cite{holland_dataset_2018}. Despite its promises, the use of AI to expand human analytic capacity can also create new forms of labor, some of which can be institutionalized (e.g., humans in charge of determining the veracity of LLM results in the face of potential hallucinations ~\cite{fox_generative_2024}). However, with the introduction of complex and often obscure LLMs---embedding built-in chains of prompts and assuming categories for dataset attributes~\cite{giner-miguelez_datadoc_2023}---many additional forms of labor might remain hidden, complicating users’ work in unanticipated ways. 

\subsection{The Blurry Views of Dataset Documentation Tools’ Broader Integration}
\label{sec:findings-integration}

Our analysis of tools in our corpus sheds light on a tendency to discuss the work of integrating documentation tools within socio-technical systems in blurry terms, often sustained on assumptions over what users would/could do with the tools, and therefore placed in a future over which authors cannot offer more than recommendations. Exceptions to this include \cite{mcmillan-major_data_2024,Bhardwaj_Gujral_Wu_Zogheib_Maharaj_Becker_2024}, who included concrete examples of integration in connection to the IRB system and the ML conference infrastructure, respectively. We also found that the discourses that creators had about what successful integration might entail and the assumptions that feed these discourses, were not always grounded in empirical data. With this analysis, we echo the results of Heger et al. study with data practitioners that foregrounded the need for “processes by which data documentation is created to be integrated with the databases, cloud platforms, and analysis tools that they use”~\cite{heger_understanding_2022}. 

\subsubsection{Discourses about Integration Work: Relying on Unsupported Assumptions}
Several creators in our corpus recognized the importance for documentation tools to be fully integrated into concrete contexts, often expressing that \textit{“[robust documentation’s] durability is enabled by sufficient support, integration, and flexibility of documentation practices”}~\cite{chmielinskiDatasetNutritionLabel2022a}. As such, they discussed plans to help users’ and adopters’ deploy and use the tools they proposed. Some were thorough in the way use cases could take place and, along with advancing a tool, provided users with a conceptualization of the documentation process. Gebru et al., for example, detailed seven steps for the documentation process all the way from the moment a dataset is motivated to when a dataset is released and requires maintenance~\cite{gebru_datasheets_2021}. While this level of detail can be helpful for future users, it may fall short in capturing the work required to successfully integrate these tools across a large variety of contexts, some of which may fall outside of this characterization, a challenge acknowledged by some authors~\cite{mcmillan-major_reusable_2021,gebru_datasheets_2021,holland_dataset_2018}. As we reported in section \ref{sec:findings:descriptive-stats}, this lack of attention to integration was a common phenomenon across a large portion of tools in our sample.
 
At the other end of the spectrum were creators who conceptualized integration efforts in terms of their perceived benefits. In many cases this conceptualization led creators to prioritize ease of use, technical features, and connection to technical features they believed would be of value to users without acknowledging existing findings in other fields that stressed other complex factors~\cite{marandi_datasheets_2025,akhtarCroissantMetadataFormat2024,petersenDataMaidYourAssistant2019,fox_generative_2024,giner-miguelez_datadoc_2023}. The tool that ~\cite{giner-miguelez_describeml_2022} created exemplifies this view of integration: \textit{“One of the main goals when designing our tool has been to keep it simple and easy to use. Therefore, the installation process is as simple as searching the plugin in the VSCode extension tab, download and enable it”}~\cite{giner-miguelez_describeml_2022}. However, this tool assumed practitioners operated within a straightforward data lifecycle model, something that research has already advised against ~\cite{Deng2023, Deng_Nagireddy_Lee_Singh_Wu_Holstein_Zhu_2022, Madaio_Chen_Wallach_Wortman_Vaughan_2024}.

Another form of working towards integration in terms of perceived benefit was to provide examples of their proposals in use~\cite{gebru_datasheets_2021,mcmillan-major_data_2023,romanOpenDatasheetsMachinereadable2024a,Data_Cards_Pushkarna_Zaldivar_Kjartansson_2022,chmielinskiDatasetNutritionLabel2022a,afzal_data_2021}, and ready-to-use templates~\cite{hutchinson_towards_2021}. These efforts mostly focused on outcomes (e.g., exemplifying the resulting artifact of using a given tool), and provided much needed guidance for potential users. Yet, few proposals took concrete steps to facilitate said integration in detail, often disregarding aspects that are critical for realistic integration such as the data governance structures from where datasets emerged~\cite{holland_dataset_2018,gebru_datasheets_2021,bender_data_2018,hutchinson_towards_2021}. 

Authors also conceptualized integration efforts in terms of perceived benefits by seeking to extend existing tools into particular fields of knowledge. A common example of this was the extension of Datasheets into the arts~\cite{srinivasan_artsheets_2021}, healthcare~\cite{siddik_datasheets_2025}, and energy sectors~\cite{heintz_datasheets_2023}, for example. This strategy, other authors argued, could lead to overlaps and/or duplication, adding to the labor and time concerns we mentioned in previous sections~\cite{afzal_data_2021}. These instances, while concrete and scoped, did not get into details about practical integration.   

Overall, our data suggests that perspectives of integration based on perceived benefits were not grounded on empirical data and rather expressed as assumptions: \textit{“…by harmonising the differing perspectives of data scientists and legal experts, proposed data-envelopes serve as a bridge between technical and legal frameworks, facilitating a more ethical and legally compliant use of historical data”}~\cite{luthra_data-envelopes_2024}

\subsubsection{Authors' Beliefs on the Integration Process}
Our analysis also suggest that authors hold at least two critical beliefs about how tools created as part of research can be integrated into realistic use contexts: integration is an iterative, long-term process, with its final success placed in an unknown future and integration takes place organically. In both cases, creators’ discourses suggest a belief that taking on the work of integration is someone else responsibility. 

The belief that integration will happen at some point in the future often emerged as a recognition of the challenges related to it: \textit{“Since principles are often compelling in theory but challenging to realize in practice, our hope is that by discussing such tradeoffs, CLeAR can serve as a realistic guide for creating documentation while considering and understanding some of the choices that will need to be made”}~\cite{The_CLeAR_Documentation_Framework_for_AI_Transparency_2024,hutchinson_towards_2021}. This focus on “future integration” relies in the role of “champions”, individuals or groups of users who can take on the labor of integrating these tools, departing from non-prescriptive recommendations offered by authors and transforming them into well-established organizational best practices~\cite{holland_dataset_2018,gebru_datasheets_2021}. This matches a related trend of authors referring to integration as an iterative, long-term process, with its final success placed in an unknown future~\cite{luthra_data-envelopes_2024,giner-miguelez_describeml_2022,hutchinson_towards_2021}. Some authors did address integration more specifically, and in reference to current structures that could aid in integrating tools: \textit{“To improve ‘findability’, we urge NeurIPS to require datasets to have metadata (not just data) assigned a persistent identifier and hosted in a searchable repository (such as Zenodo)”}~\cite{Bhardwaj_Gujral_Wu_Zogheib_Maharaj_Becker_2024,akhtarCroissantMetadataFormat2024}. This kind of effort then paves the way for studies looking at the use of these integrated tools across dataset repositories, for example. ~\citeauthor{Yang_Liang_Zou_2023} study of the use of Data and Model cards across the Hugging Face repository are one recent example.

As we found, authors also framed integration as an organic process assuming enthusiastic users willing to discuss the ideas advanced by authors and turn them into successful practices: \textit{“Responses to documentation questions are not intended to be prescriptive, nor are they completely comprehensive. Instead, they should be considered as one of many valid responses to this line of inquiry, and as a way to provoke further thought and discussion.”}~\cite{diaz_crowdworksheets_2022}. Once again, this adds to the labor of documentation often inadvertently placed in users often complicating the adoption of these tools. 

\section{Discussion}
\label{sec:discussion}

Our exploration highlighted that, with few exceptions (e.g., \cite{mcmillan-majorLanguageDatasetDocumentation2023, boyd_datasheets_2021}), most authors in the corpus conceptualized dataset documentation tools in ways that may hinder their adoption and standardization. These views clustered around four themes: (1) a disagreement on how to operationalize the value of documentation; (2) a tendency to design decontextualized tools disconnected from user experiences; (3) an increase of undiscussed labor demands, even when moving towards automation; and (4) a view of tool integration as an aspirational future endeavor. These patterns echo findings from previous work exploring the design and adoption of other forms of Responsible AI (RAI) tools (e.g.,\cite{Wong_Madaio_Merrill_2023, Madaio_Chen_Wallach_Wortman_Vaughan_2024, Balayn_Yurrita_Yang_Gadiraju_2023, Deng_Nagireddy_Lee_Singh_Wu_Holstein_Zhu_2022, madaio_2022}), suggesting a broader, systemic misalignment between how industry and academics understand RAI work. 
In what follows, we discuss how our findings shed light on this misalignment. Further, we argue that the HCI community interested in supporting RAI initiatives needs to pause the focus on designing tools; continuing to produce documentation tools that offload responsibilities onto practitioners without addressing the industry-academia misalignment can be counterproductive for sustainable documentation practices. Building on the work of \cite{Wong_Madaio_Merrill_2023, Madaio_Chen_Wallach_Wortman_Vaughan_2024, Deng_Nagireddy_Lee_Singh_Wu_Holstein_Zhu_2022, vera_2019, Wong_2021_tactics, ali_2023} and outlier cases in our corpus \cite{akhtarCroissantMetadataFormat2024, piorkowskiFieldStudyHumanCentered2024}, we call on the HCI community to lead a shift from a design mindset to a research mindset that centers on systemic issues preventing academia from responding to industry’s documentation goals. To that end, we propose an HCI research agenda that supports industry and academia in developing sustainable dataset documentation practices. 

\subsection{Dataset Documentation Tools and the Systemic Industry-Academia Misalignment: The Need to Rethink the Focus on Design} 
\label{sec:discussion:misalignment}

Our findings highlight that corpus authors, mostly academics (n=46) from CS and ML fields, did not work closely with their end-users or tailor to their specific contexts. Prior research shows that industry environments often operate under fast-paced timelines governed by launches of new software products \cite{madaio_2022, ali_2023}, which can hinder companies’ ability to share contextual knowledge with academic researchers ~\cite{BrookingInstitute2023, Eastwood_2023a, ahmed_2023}. Our findings illuminate three key factors shaping the misalignment between academic conceptualizations of documentation and industry needs: a lack of industry-validated consensus on documentation, academia’s disengagement with industry, and academia’s disconnection with HCI knowledge. Furthermore, we argue that, given academia’s response to these factors, it is crucial to pause the development of additional documentation tools. Instead, we must address the systemic factors that currently hinder the standardization and adoption of documentation practices in the industry. 

\subsubsection{Non-Standardized, Contrasting Views on Documentation: A Lack of Industry-Validated Consensus}

As we observed in our corpus, most authors pursued transparency and accountability through non-standardized and sometimes contrasting interpretations. To articulate their perspectives, many drew inspiration from tools and processes in the fields of Nutrition, Engineering, and Software Engineering, among others. While some corpus authors did study their tools’ ability to prompt reflection \cite{Data_Cards_Pushkarna_Zaldivar_Kjartansson_2022, boyd_datasheets_2021}, the majority provided little empirical evidence that their tools could achieve their intended goals. This pattern aligns with prior research showing that RAI tools often lack clear definitions of RAI concepts (e.g., ethical values, social impact, equitable representation)~\cite{ali_2023, Polonski2018}. Operating under multiple, non-standardized understandings of ethics, however, can be problematic for practitioners: it forces them to engage in definitional work with no success criteria and the pressure of organizational leadership to prioritize business discourses and interests~\cite{ali_2023, Polonski2018}.

The tendency of tools in our corpus---and RAI tools in general---to offer diverse perspectives on critical ethical concepts (e.g., the value of documentation) suggests that, in the absence of industry-provided clarity, academics resort to connections with theoretical concepts from different disciplines without the obligation to ground them in practice. To address this problem, prior research has recommended designing tools that help practitioners learn and embrace non-technical dimensions of RAI work~\cite{Wong_Madaio_Merrill_2023, Balayn_Yurrita_Yang_Gadiraju_2023} with the goal of bridging disciplinary and organizational divides~\cite{Wong_Madaio_Merrill_2023, Deng_Nagireddy_Lee_Singh_Wu_Holstein_Zhu_2022}. However, our findings indicate that, in the case of documentation tools, such recommendations may inadvertently add to practitioners’ struggles. If the industry fails to reach consensus on the value of documentation, academia will continue to produce contrasting perspectives, further complicating practitioners’ decision-making processes.

\subsubsection{Decontextualized Labor Demands: An Academic Disengagement with Industry}

Our findings highlight that corpus authors tended not to engage with end users. Despite some exceptional cases, such as~\cite{piorkowskiFieldStudyHumanCentered2024, romanOpenDatasheetsMachinereadable2024a}, in which authors involved key stakeholders in some part of the creation process (n=10) (see Figure~\ref{fig_3_stakeholder_integration_sankey}), most failed to define an end-user profile, and championed a context-agnostic approach to documentation. As a result, tools in our corpus tended to demand new, undiscussed forms of labor from practitioners such as tools’ contextualization and the assessment of potential cultural harms. As~\citeauthor{heger_understanding_2022} and ~\citeauthor{winecoff2025} have emphasized, ethical interpretation of dataset work can be particularly hard for practitioners: in contrast with data interpretation work, ethical interpretations require a type of expertise that most technical practitioners do not have and knowledge of contexts they are often not familiar with~\cite{heger_understanding_2022,winecoff2025}. While automated documentation tools purported to reduce labor, they could further obscure ethical interpretative labor by driving practitioners to engage in activities where the potential ethical risks might cause a cascading effect and thus be more complex to identify ~\cite{winecoff2025}~(e.g., deciding which datasets are uninteresting enough to exclude).

Previous work analyzing the design of RAI toolkits found a similar tendency: these toolkits often operated under a “decontextualized approach to ethics” \cite{Wong_Madaio_Merrill_2023} that disregarded the various human and organizational factors shaping RAI work, and hindered adoption by placing intensive contextualization labor on practitioners \cite{Wong_Madaio_Merrill_2023, Madaio_Chen_Wallach_Wortman_Vaughan_2024, Balayn_Yurrita_Yang_Gadiraju_2023, Deng_Nagireddy_Lee_Singh_Wu_Holstein_Zhu_2022}. The recurrent decision from tools’ creators to produce decontextualized RAI tools suggests more pathways to engage with industry experiences may be needed. To redress the negative impact that a “decontextualized approach to ethics” has on ethical interpretation demands, existing research has proposed designing for moments of positive ambiguity~\cite{Madaio_Chen_Wallach_Wortman_Vaughan_2024} that prompt practitioners to realize they need to reflect on ethical issues, discuss them with others, and reach a consensus. Our findings indicate, however, that this design recommendation might be dangerous in the context of documentation tools: if the academics creating tools continue to operate with such distance from industry, they will not be able to identify and evaluate productive moments for positive ambiguity and could, instead, end up creating more undiscussed, decontextualized labor demands.

\subsubsection{Blurry Integration Possibilities: An Academic Disconnection with HCI Knowledge}

Finally, most of the tools in our corpus described possibilities for tool integration in blurry terms (with some notable exceptions, such as~\cite{Fact_Sheets_Arnold_Bellamy_Hind_Houde_Mehta_Mojsilović_Nair_Ramamurthy_Olteanu_Piorkowski_2019, Data_Cards_Pushkarna_Zaldivar_Kjartansson_2022, gebru_datasheets_2021}). This further highlights two characteristics of how academia is handling the academia-industry misalignment in the context of documentation tools. First, we observed a tendency among corpus authors to release the initial versions of tools without evaluation evidence or detailed integration guidance. Such a tendency suggests that, in the face of a distance from industry, academics resort to a practice of generating knowledge through small iterations that “somebody”---another researcher or industry actor---might eventually use or further extend. However, as previous work on RAI tools has found, \textit{“[RAI] tools and practices that do not align with practitioners’ workflows and organizational incentives may not be used as intended or even used at all”} \cite{lee_2021, Madaio2020, Rakova2021, richardson_2021}.

Second, our findings suggest that a stronger connection with HCI research on RAI practices could have helped corpus authors propose more realistic integration guidance. As our previous discussion sections stressed, the majority of authors disregarded existing HCI research on RAI practices and repeated trends that have been already found as counterproductive for RAI goals~\cite{wang_2023, sadek_2024}. Further, with a few exceptions (e.g., as~\cite{piorkowskiFieldStudyHumanCentered2024, boyd_datasheets_2021}), tools did not resort to human-centered research knowledge and methods to enhance technology integration. Against this backdrop, expecting tools’ authors to follow design alternatives from HCI researchers guiding integration might be unrealistic. For example, HCI-related research has already suggested that a productive integration pathway is to design for challenging linear views of data lifecycles~\cite{Deng2023, Deng_Nagireddy_Lee_Singh_Wu_Holstein_Zhu_2022, Madaio_Chen_Wallach_Wortman_Vaughan_2024}. Many tools in our corpus, however, assumed practitioners operated within a straightforward data lifecycle model~\cite{giner-miguelez_datadoc_2023, giner-miguelez_describeml_2022, fabris_tackling_2022, diaz_crowdworksheets_2022}.

\subsection{Towards Sustainable Dataset Documentation Practices: An HCI Research Agenda}

Our findings highlight how the misalignment between industry and academic conceptualizations of documentation tools---mainly coming from CS and ML fields---produces documentation tools that may devalue dataset work, thus hindering the standardization and adoption of documentation practices. To reverse the devaluation of dataset work, Hutchinson et al. called for a radical cultural shift involving the “broader systems and ecologies in which datasets are maintained and used” \cite{hutchinson_towards_2021}. We draw inspiration from their work and propose that, to change how documentation tools are conceptualized, we need to shift from a design to a research perspective that identifies strategies for addressing the systemic factors that shape the industry-academia misalignment. Furthermore, we argue that the field of HCI, which has developed multidisciplinary and critical approaches for democratizing the design of technologies, including RAI tools \cite{sadek_2024, wang_2023, Madaio2020, Deng_Guo_Devrio_Shen_Eslami_Holstein_2023}, is best positioned to lead this shift. Next, we discuss a research agenda for HCI scholars working towards effective and sustainable documentation practices.

\subsubsection{Strategies and Methods for an Industry-Consensuated View of Documentation}

As our findings highlight, the lack of industry-validated consensus on the value of documentation has led academia to generate numerous, contrasting proposals for how documentation can be valuable. Further, as we observed, the distance between academics who conceptualize tools and actual industry practice prevents proposals from measuring effectiveness in ways that matter to industry. Reaching industry-validated consensus on ethics-related concepts, however, is not seamless. As \cite{Balayn_Yurrita_Yang_Gadiraju_2023, Wong_2021_tactics, ali_2023} have stressed, power dynamics complicate this task: organizational leadership often uses its power to prioritize AI perspectives that align with business directives, which fragments and diverts practitioners’ efforts away from AI responsibility and ethics \cite{Rakova2021, Deng2023}. As such, to prevent the creation of non-standardized, contrasting views of documentation that force practitioners to engage in ethics definitional work, it is critical to identify the best ways to align the interests of all actors around documentation.

The field of HCI can help identify consensus strategies that resist organizational leadership’s efforts to “tame” ethics \cite{Metcalf2019OwningEthics, Phan2021EconomiesOfVirtue, Phan2022EconomiesOfVirtue}. For example, HCI researchers could unpack the difficulties and successes behind the creation of domain-specific documentation tools like the few ones in our corpus that attempted to propose disciplinary consensus on concepts championed by generic tools. Similarly, HCI researchers could study initiatives where professional or non-profit organizations have worked to standardize and disseminate RAI practices and identify aspects such as stakeholders’ interests, resources used during the consensus-building process, unavailable resources, and mechanisms to resist subordinating ethical debates to corporate interests. Documenting these aspects could help RAI researchers, practitioners, and organizations define, with the industry, how documentation could be most effective.

Practitioners working within the industry may benefit from strategies to collectively work towards consensus on RAI concepts while navigating internal power dynamics ~\cite{ali_2023, Polonski2018, Balayn_Yurrita_Yang_Gadiraju_2023, Wong_2021_tactics}. While individual professionals often employ useful “soft” resistance tactics to work towards more “values-conscious” ends, as ~\citeauthor{Wong_2021_tactics} and ~\citeauthor{ali_2023} explained, relying on individuals rather than collective action to promote change can be detrimental for workers, especially to those from marginalized backgrounds and positions of relatively low power within their companies. Workers do not always have “the agency and authority to make or contest values and ethics decisions;”\cite{Wong_2021_tactics} their acts of resistance might result in the company firing the practitioner or re-assigning them to a different task. Further, individual resistance strategies are emotionally demanding and vulnerable to subversion from “the dominant discourses and logics of the technology industry” \cite{Wong_2021_tactics, Metcalf2019OwningEthics, Madaio_Chen_Wallach_Wortman_Vaughan_2024}. The field of HCI can intervene by illuminating methods to support practitioners’ coming together in communities of practice with other workers. To minimize the production of tools championing contrasting views on the value of documentation and adding undiscussed forms of labor, HCI could explore collaborative and participatory methods for helping practitioners: (1) explore definitions on documentation and desirable documentation practices, (2) develop strategies to document ethical issues, (3) empirically explore the role of automation in documentation, (4) create industry-valued metrics to demonstrate the effectiveness of documentation, and (5) establish effective practices for communicating and negotiating changes with organizational leadership.

\subsubsection{Human-Centered Knowledge and Methods for Conceptualizing Documentation Tools in Industry Contexts}

Our findings showed how the majority of the documentation tools in our corpus stemmed from academics in the fields of CS and ML. These academics, as we have already discussed, tended to operate in disconnection with HCI knowledge, both in relation to RAI practices and human-centered methodologies for understanding users and their context. As a result, they leveraged the freedom of academia to experiment with theories, producing context-agnostic tools based on academics’ own experiences, and with no clarity on who potentially relevant stakeholders would be. Furthermore, they presented initial iterations of their proposals with vague, aspirational integration mechanisms (e.g., ease of use, publication of templates or examples). To ensure that documentation tools do not add decontextualized forms of labor and factor concrete integration strategies, it is critical that academics develop a closer connection with HCI knowledge.

To achieve this connection, the field of HCI can explore with creators the reasons behind the disconnection, including possible communication aspects hindering the application of these findings into their designs. Further, HCI researchers could work with authors who emphasized the importance of generating evidence and metrics about the impact of documentation tools (e.g., \cite{Data_Cards_Pushkarna_Zaldivar_Kjartansson_2022, boyd_datasheets_2021} in exploring their experiences, and disseminating this knowledge to other tool creators. The overall goal is to identify barriers and opportunities in ensuring that tool creators can utilize evidence-based research on the value of documentation. Finally, HCI researchers can engage in participatory design activities with tool creators to envision better ways to present, disseminate, and operationalize documentation-related findings, including evidence of documentation effectiveness in supporting RAI goals.  

The efforts of some authors in our corpus to approach end-users to understand, design with, or gather feedback, emphasizes the difficulties tool creators may face when reaching out to industry practitioners. As our findings show, some of these authors resorted to online methods to gather practitioners’ feedback without compromising their identities (e.g., \cite{gebru_datasheets_2021}.) These efforts, however, were not common in our corpus, which suggests that tool authors need support not only in learning more about possible HCI methods, but also in considering how to tailor them for the specific needs and level of distance from industry actors. To that end, HCI researchers can work with creators to identify the obstacles they foresee in working closely with practitioners and understanding their workflows. Furthermore, HCI researchers can explore appropriate participatory design methodologies to collaborate with creators in developing user study methodologies that help them navigate these obstacles, enabling the analysis of interest alignment and prioritizing collective power.

\section{Limitations}
In this study, we take an interpretative approach to data analysis that focuses on the motivations, conceptualizations, and execution of data documentation proposals. We recognize that this decision leaves open the possibility of additional findings focused on other themes, especially descriptive comparisons across tools. We also do not elicit practitioners’ views first-hand in the context of this study. Instead, we lean on empirical scholarship in HCI, foregrounding prior work and connecting its results to our findings.
Our study focuses on a subset of tools related to dataset documentation, specifically those about which information is publicly available in academic databases and on the open web. As such, private resources, tools or frameworks used by organizations in the private sector are outside of our domain of analysis. 
We argue that increasing focus on producing empirical evidence of how documentation meaningfully contributes to increasing accountability and transparency, can help practitioners better judge the value of tools for their particular applications. We acknowledge that some of this work intersects with studies around AI auditing and recent studies focused on practitioners workflows; this intersection falls outside of the scope of this paper.

\section{Conclusion}
\label{sec:conclusion}

Data transparency and accountability remain important foundations for the responsible development of automated systems. Appropriate dataset documentation is a key tenet of these pillars. Despite efforts to facilitate dataset documentation through tools including software applications and frameworks, standardized adoption and integration of these tools remains scarce. To reveal barriers and opportunities towards broader use of these tools, we conducted a scoping review examining the motivations and conceptualizations advanced by designers of dataset documentation tools. Our contributions are two-fold. First, we foreground aspects of these tools significantly impacting their adoption, including a spectrum of perspectives on documentation practices, a devaluing of the dataset infrastructure, an unquestioned transition towards automation, and an aspirational view around the integration of these tools. Second, we propose a radical shift in how to design Responsible AI tools focusing on institutions rather than individuals. We discuss critical actions the HCI community can take to support this shift.

\section*{Acknowledgements}
We would like to thank Jenny Barry, Maximillian Clee, Jessica Hodgins, and the conference referees for their helpful comments.

\bibliographystyle{unsrtnat}
\bibliography{references}  %%% Uncomment this line and comment out the ``thebibliography'' section below to use the external .bib file (using bibtex) .

\appendix
\clearpage
\section{Further Descriptive Analyses}
\label{sec:appendix-descriptive-analyses}

\begin{figure}[h]
    \centering
    \includegraphics[alt={Figure 4 is a stacked vertical bar graph showing the count of documentation tools by degree of automation from 2010 to 2025. The vertical axis shows Count, ranging from 0 to 14. The horizontal axis shows Year. The total number of tools increases significantly over time, from 1 tool in 2010 to a peak of 13 tools in 2024, before declining to 3 in 2025. Hybrid tools dominate most years, with Manual tools increasing notably in 2021-2024. The data are summarized in the following table: Count of Documentation Tools by Degree of Automation and Year. Year 2010: Manual 0, Hybrid 1, Automated 0, Total 1. Year 2016: Manual 0, Hybrid 1, Automated 0, Total 1. Year 2018: Manual 1, Hybrid 1, Automated 0, Total 2. Year 2019: Manual 1, Hybrid 1, Automated 2, Total 4. Year 2020: Manual 2, Hybrid 1, Automated 0, Total 3. Year 2021: Manual 0, Hybrid 6, Automated 0, Total 6. Year 2022: Manual 4, Hybrid 5, Automated 1, Total 10. Year 2023: Manual 0, Hybrid 7, Automated 1, Total 8. Year 2024: Manual 5, Hybrid 5, Automated 3, Total 13. Year 2025: Manual 1, Hybrid 1, Automated 1, Total 3.}, width=0.9\linewidth, height=6cm]{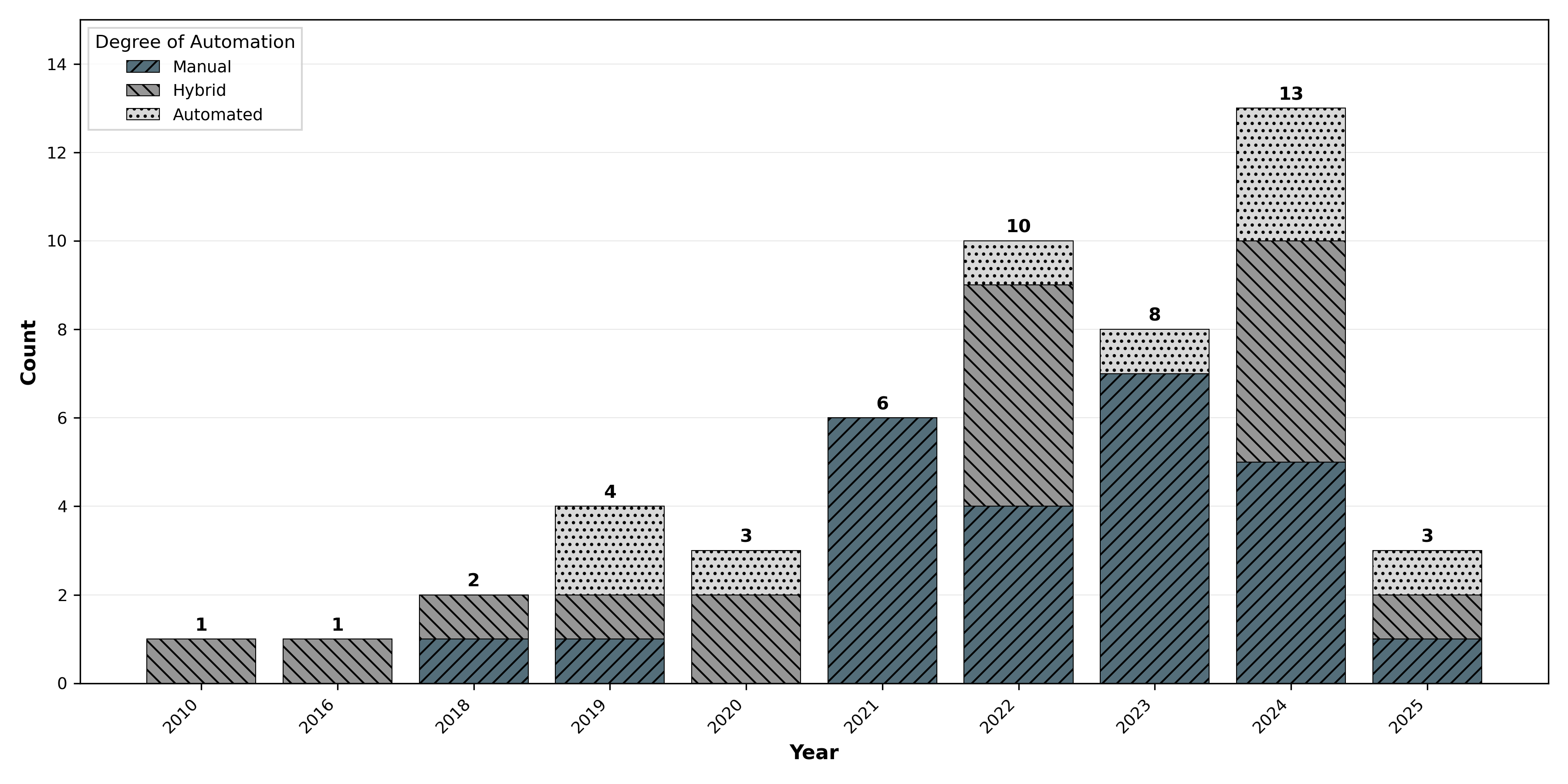}
    \caption{Distribution of approaches to the construction of dataset documentation over the years. Data shows a prevalence of manual approaches with a sustained increase in automated and semi-automated approaches between 2022 and 2025.}
    \label{fig_4_distribution_of_automation_over_time}
\end{figure}

\begin{figure}[h]
    \centering
    \includegraphics[alt={Figure 5 contains two grouped vertical bar graphs showing Tool Evaluation and Tool Integration by type of tool. Each graph has Count on the vertical axis ranging from 0 to 14, and Type of tool on the horizontal axis with six categories: toolkit, framework, application, datasheet, markup format, and schema. Each category shows two bars: Yes and No. The data are summarized in the following tables. Tool Evaluation: toolkit Yes 0, No 2; framework Yes 2, No 14; application Yes 4, No 9; datasheet Yes 4, No 9; markup format Yes 2, No 3; schema Yes 0, No 2. Most tools across all types have not been evaluated. Tool Integration: toolkit Yes 0, No 2; framework Yes 3, No 13; application Yes 5, No 8; datasheet Yes 2, No 11; markup format Yes 3, No 2; schema Yes 0, No 2. Most tools across all types have not been integrated, with application showing the highest count of integration at 5 tools.}, width=0.9\linewidth, height=6cm]{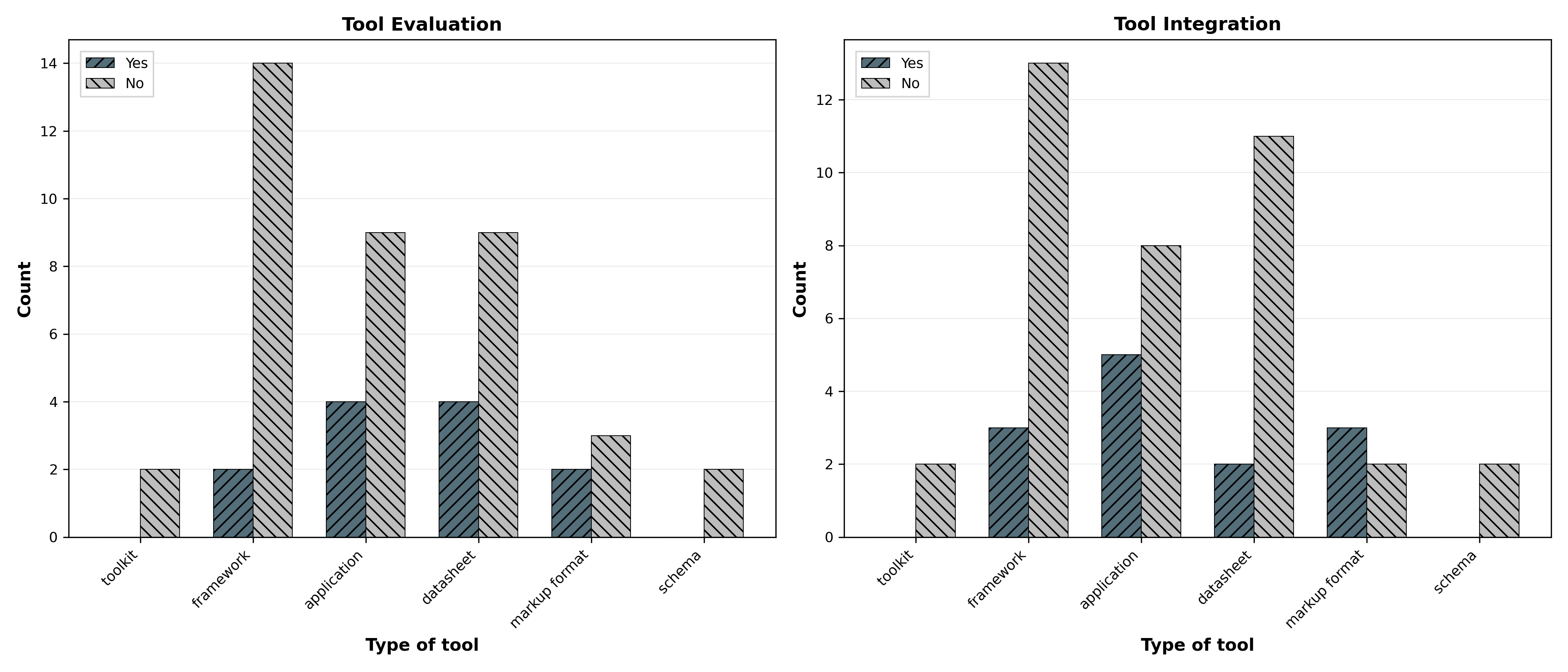}
    \caption{Comparative analysis of different types of tools that included tool evaluation or features towards integration. \textbf{Left:} Count of proposals that included an evaluation study at any stage of the design or integration of the tool. \textbf{Right:} Percentage of proposals that included stakeholders during the design/testing stages compared to the number of proposals that included concrete features or guidance towards integration to stakeholders.}
    \label{fig_5_evaluation_and_integration_across_tools}
\end{figure}

\begin{figure}[ht]
    \centering
    \includegraphics[alt={Figure 6 is a heatmap showing the count of documentation tools by audience type and tool type. The horizontal axis shows Type of tool with six categories: application, datasheet, framework, markup format, schema, and toolkit. The vertical axis shows Audience with nine categories: data curators, data experts, data organizations, data practitioners, dataset auditors, dataset creators, dataset practitioners, dataset researchers, and dataset users. Color intensity indicates count, ranging from 0 (light) to 14 (dark). Dataset creators show notably higher counts across most tool types compared to other audiences. The data are summarized in the following table. Data curators: application 1, datasheet 1, framework 0, markup format 2, schema 0, toolkit 0. Data experts: application 0, datasheet 3, framework 2, markup format 0, schema 0, toolkit 0. Data organizations: application 0, datasheet 0, framework 0, markup format 0, schema 1, toolkit 0. Data practitioners: application 0, datasheet 1, framework 1, markup format 0, schema 0, toolkit 0. Dataset auditors: application 1, datasheet 1, framework 0, markup format 1, schema 0, toolkit 0. Dataset creators: application 12, datasheet 13, framework 14, markup format 3, schema 2, toolkit 2. Dataset practitioners: application 1, datasheet 0, framework 3, markup format 1, schema 0, toolkit 0. Dataset researchers: application 0, datasheet 0, framework 1, markup format 1, schema 0, toolkit 0. Dataset users: application 3, datasheet 5, framework 3, markup format 1, schema 0, toolkit 2.}, width=\linewidth]{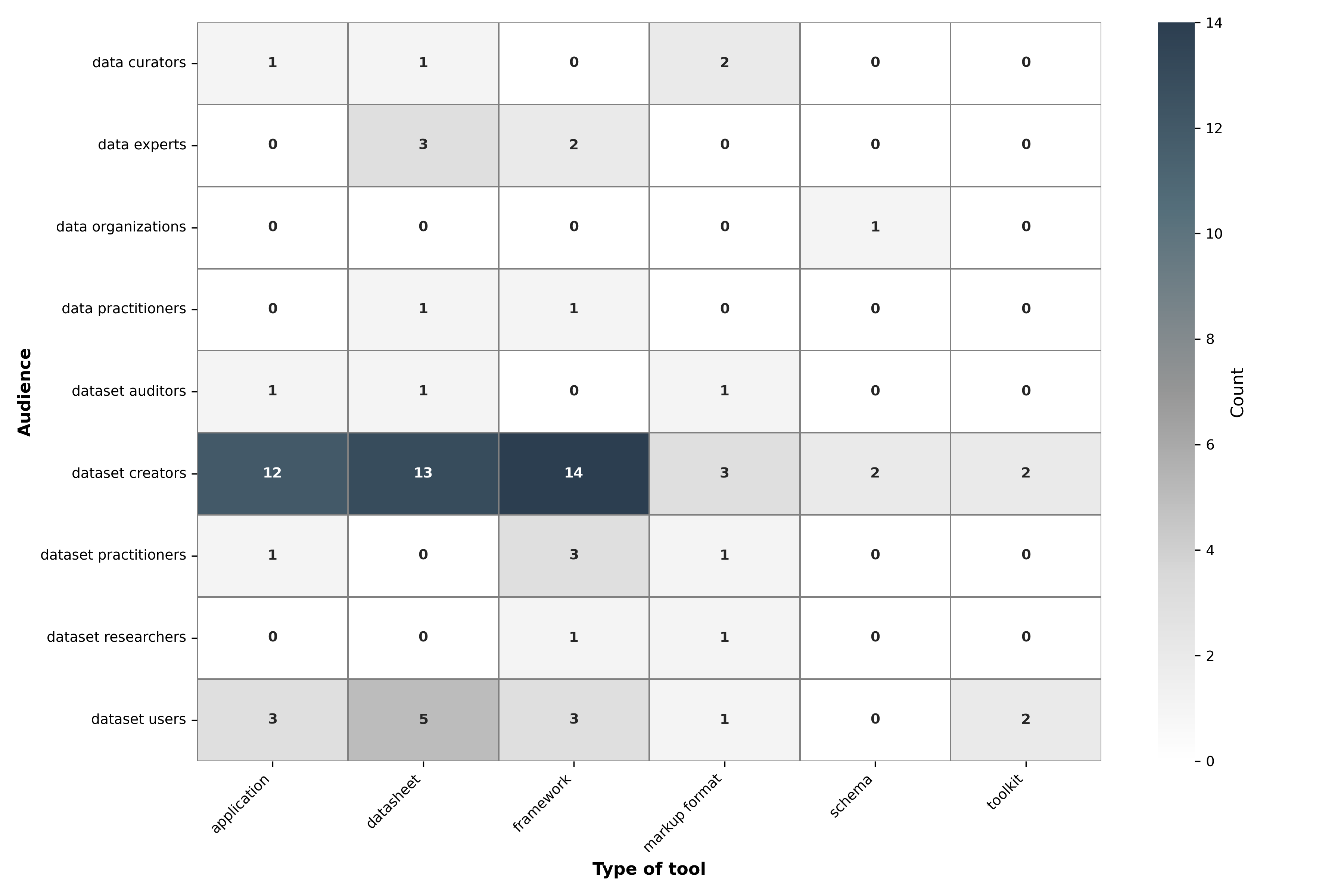}
    \caption{Overlap of different types of tools across different audiences as stated by authors. Data suggests ‘dataset creators’ as a main category of interest for authors, and ‘applications, datasheets and frameworks, as the most preferred categories of tools.}
    \label{fig_6_distribution_tool_type_across_audiences}
\end{figure}

\clearpage
\section{Data sourcing and screening process}
\label{sec:appendix-prisma-figure}

\begin{figure}[h]
    \centering
    \vspace{-0.8em} 
    \includegraphics[alt={Figure 7 is a PRISMA flow diagram showing the systematic literature review process across four stages: Identification, Screening, Eligibility, and Included. Identification stage: Articles identified based on queries (n=1491) from ACM DL (n=332), IEEE Xplore (n=891), ScienceDirect (n=228), ArXiv (n=32), NeurIPS (n=3), AAAI (n=0), and ACL Anthology (n=5). Articles identified from other sources (n=150) from Google Search (n=41), Google Scholar (n=77), and Snowballing (n=32). Screening stage: Records after duplicates removed (n=1582). Records screened by Title/Abstract (n=1582). Articles excluded (n=1408) from ACM DL (n=286), IEEE Xplore (n=816), ScienceDirect (n=211), ArXiv (n=21), NeurIPS (n=3), ACL Anthology (n=0), Google Scholar (n=60), Google Search (n=11), and Snowballing (n=0). Eligibility stage: Full text articles assessed for eligibility (n=174). Full text articles excluded (n=115) from ACM DL (n=28), IEEE Xplore (n=25), ScienceDirect (n=13), ArXiv (n=7), NeurIPS (n=2), ACL Anthology (n=3), Google Scholar (n=9), Google Search (n=11), and Snowballing (n=17). Included stage: Full text articles (n=59) from ACM DL (n=18), IEEE Xplore (n=3), ScienceDirect (n=4), ArXiv (n=5), NeurIPS (n=1), ACL Anthology (n=2), Google Scholar (n=8), Google Search (n=3), and Snowballing (n=15).}, width=0.85\linewidth]{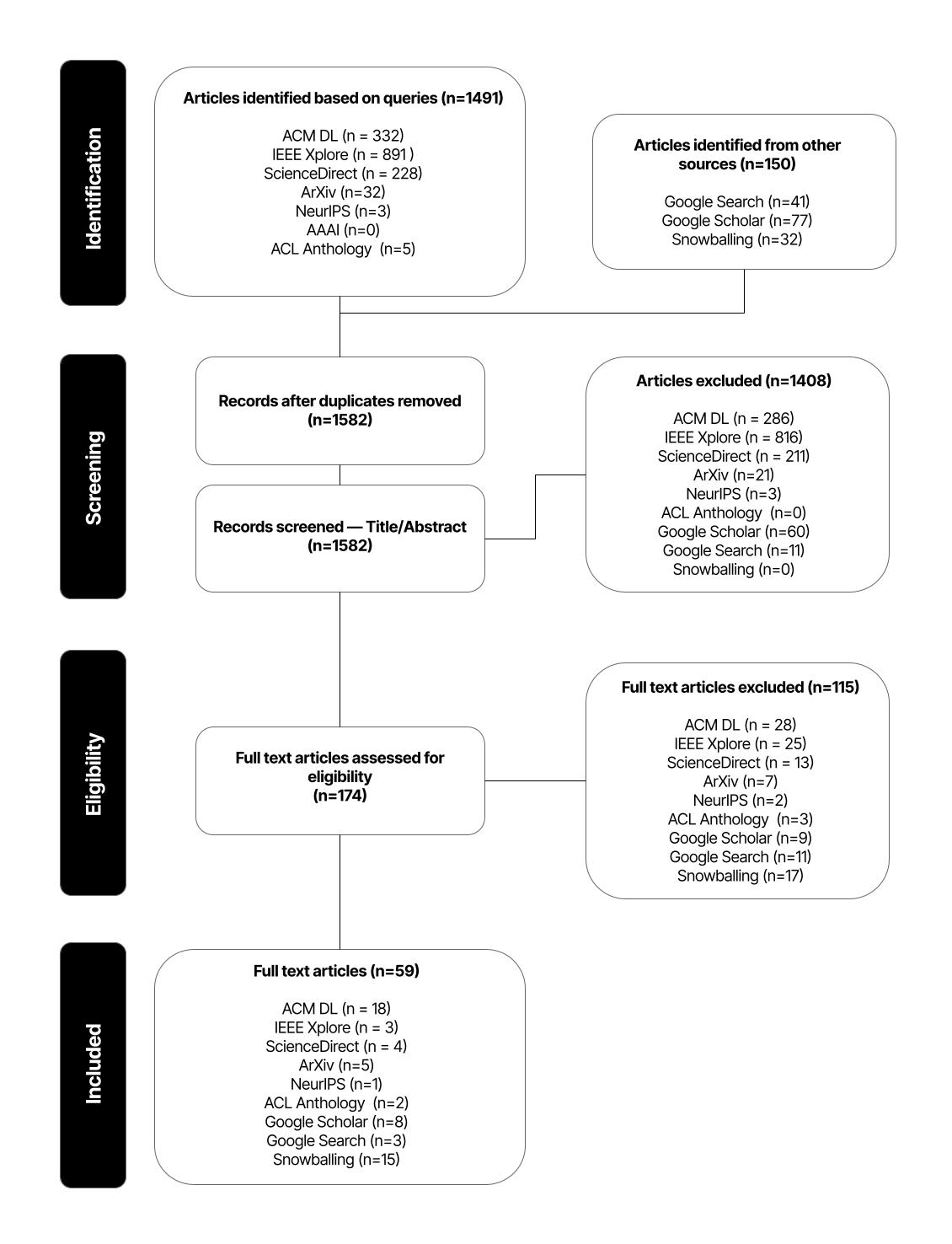}
    \vspace{-0.8em} 
    \caption{Adapted flowchart from PRISMA outline, describing demonstrating how we searched, screened, and selected items for inclusion in our corpus}
    \label{fig_7_prisma_datasets}
\end{figure}

\clearpage
\onecolumn
\section{Data collection search queries}
\label{sec:appendix-search-queries}

%\FloatBarrier
\begin{small}
    \begin{longtable}[*]{p{1.6cm}p{5.5cm}p{1.5cm}p{2.0cm}p{2.5cm}}
        %\renewcommand{\arraystretch}{1.5}
        %\begin{tabular}{m{1.5cm}m{6.0cm}m{1.5cm}m{2.0cm}m{2.0cm}}
        %\toprule
        \hline
        \rowcolor{black}
        \multicolumn{5}{c}{\color{white}\textbf{Keywords Queries}} \\ \hline
        \multicolumn{1}{c}{\textbf{Resource}} &
        \multicolumn{1}{c}{\textbf{Query}} &
        \multicolumn{1}{c}{\textbf{Date}} &
        \multicolumn{1}{c}{\textbf{Field}} &
        \multicolumn{1}{c}{\textbf{Notes}} \\ \hline
        \textbf{NeurIPS} & "dataset documentation" OR "dataset transparency" OR "dataset accountability" OR "dataset provenance" OR "dataset datasheet" OR “dataset management”  & Feb 1st & All fields search and manual search over Dataset and Benchmarks repositories & This track has contributions from 2021-2024
        \\ \hline
        \textbf{ACM} & Title:(``dataset'' AND (``documentation'' OR ``transparency'' OR ``accountability'' OR ``provenance'' OR ``datasheet'' OR ``management'')) OR Abstract:(``dataset'' AND (``documentation'' OR ``transparency'' OR ``accountability'' OR ``provenance'' OR ``datasheet'' OR ``management'')) OR Keyword:(``dataset'' AND (``documentation'' OR ``transparency'' OR ``accountability'' OR ``provenance'' OR ``datasheet'' OR ``management'')) & Feb 7th & title-abstract-keywords &   
        \\ \hline
        \textbf{IEEE} & (``Document Title'':``dataset'' OR ``Abstract'':``dataset'' OR ``Author Keywords'':``dataset'') AND (``Document Title'':``documentation'' OR ``Document Title'':``transparency'' OR ``Document Title'':``accountability'' OR ``Document Title'':``provenance'' OR ``Document Title'':``datasheet'' OR ``Document Title'':``management'' OR ``Abstract'':``documentation'' OR ``Abstract'':``transparency'' OR ``Abstract'':``accountability'' OR ``Abstract'':``provenance'' OR ``Abstract'':``datasheet'' OR ``Abstract'':``management'' OR ``Author Keywords'':``documentation'' OR ``Author Keywords'':``transparency'' OR ``Author Keywords'':``accountability'' OR ``Author Keywords'':``provenance'' OR ``Author Keywords'':``datasheet'' OR ``Author Keywords'':``management'') & Feb 7th & title-abstract-keywords & ~ \\ \hline
        \textbf{ScienceDirect} & (``dataset documentation'') OR (``dataset transparency'') OR (``dataset accountability'') OR (``dataset provenance'') OR (``dataset datasheet'') OR (``dataset management'') & Feb 7th & all fields search & ~ \\ \hline
        \textbf{ArXiv} & order: -announced\_date\_first; size: 200; classification: Computer Science (cs), Economics (econ), Electrical Engineering and Systems Science (eess), Statistics (stat); include\_cross\_list: True; terms: AND all=``dataset documentation''; OR all=``dataset transparency''; OR all=``dataset accountability''; OR all=``dataset provenance''; OR all=``dataset datasheet'' OR all=``dataset management'' & Feb 12th & all fields search & ~ \\ \hline
        \textbf{AAAI} & ``dataset documentation'' OR ``dataset transparency'' OR ``dataset accountability'' OR ``dataset provenance'' OR ``dataset datasheet'' OR ``dataset management'' & Feb 18th & all fields search & ~ \\ \hline
        \textbf{ACL Anthology} & ``dataset documentation'' OR ``dataset transparency'' OR ``dataset accountability'' OR ``dataset provenance'' OR ``dataset datasheet'' OR ``dataset management'' & Feb 21st & all-fields & Search done on Zotero over full ACL catalog BibTeX file available at https://aclanthology.org/ \\ \hline
        \textbf{Google Search} & ``dataset documentation'' OR ``dataset transparency'' OR ``dataset accountability'' OR ``dataset provenance'' OR ``dataset datasheet'' OR ``dataset management'' & Feb 24th & advanced search with omitted results included & Used the 5-pages-of-noise or data saturation rule. Done from a Cambridge, MA IP address over incognito mode on Brave browser v1.75.180. \\ \hline
        \textbf{Google Scholar} & ``dataset documentation'' OR ``dataset transparency'' OR ``dataset accountability'' OR ``dataset provenance'' OR ``dataset datasheet'' OR ``dataset management'' & Feb 24th & advanced search & Used the 5-pages-of-noise or data saturation rule. Done from a Cambridge, MA IP address over incognito mode on Brave browser v1.75.180. \\ 
            \arrayrulecolor{black}\bottomrule
        %\end{tabular}
        \noalign{\vskip 1em}
        \caption{\textcolor{black}{Detailed queries as used for each different resource. ‘Field’ denotes the search fields that were leveraged within each resource in order to return results. }}
        \label{fig:table-queries}
    \end{longtable}
\end{small}
\twocolumn
%\FloatBarrier

\clearpage
\onecolumn
\section{Dataset corpus description}
\label{sec:appendix-corpus-description}

\begin{small}
    \begin{longtable}{p{4.5cm}p{1.0cm}p{2.5cm}p{2.0cm}p{2.0cm}}
    \hline
    \rowcolor{black}
    \multicolumn{5}{c}{\color{white}\textbf{Corpus Description}} \\ \hline
    \multicolumn{1}{c}{\textbf{Tool/Item}} &
    \multicolumn{1}{c}{\textbf{Year}} &
    \multicolumn{1}{c}{\textbf{Audiences addressed}} &
    \multicolumn{1}{c}{\textbf{Degree of Automation}} &
    \multicolumn{1}{c}{\textbf{Type of Tool}} \\ \hline
    \textit{A domain-specific language for describing machine learning datasets \cite{giner-miguelez_domain-specific_2023}} & 2023 & dataset creators & Manual & Application \\ \hline
    \textit{A Field Study of a Human-Centered Process for Increasing AI Transparency \cite{piorkowskiFieldStudyHumanCentered2024}} & 2024 & N/A & N/A & Study \\ \hline
    \textit{A generative benchmark creation framework for detecting common data table versions \cite{fox_generative_2024}} & 2024 & dataset creators & Automated & Framework\\ \hline
   \textit{A Methodology for Creating AI FactSheets \cite{richards_methodology_2020}} & 2020 & dataset creators, data curators & Hybrid & Datasheet \\ \hline
   \textit{A Standardized Machine-readable Dataset Documentation Format for Responsible AI \cite{jainStandardizedMachinereadableDataset2024}} & 2024 & dataset creators, dataset users, dataset auditors & Manual & Markup format \\ \hline
    \textit{Aether Data Documentation Template \cite{Microsoft_2022_Aether}} & 2022 & dataset creators, dataset auditors & Manual & Datasheet \\ \hline
    \textit{Artsheets for Art Datasets \cite{srinivasan_artsheets_2021}} & 2021 & dataset creators, dataset users, dataset users & Manual & Datasheet \\ \hline
    \textit{Augmented datasheets for speech datasets and ethical decision-making \cite{papakyriakopoulos_augmented_2023}} & 2023 & dataset creators & Manual & Datasheet \\ \hline
    \textit{Can machines help us answering question 16 in datasheets, and in turn reflecting on inappropriate content? \cite{schramowski_can_2022}} & 2022 & dataset creators, data curators & Hybrid & Application  \\ \hline
    \textit{Clear and precise specification of ecological data management processes and dataset provenance \cite{osterweil_clear_2010}} & 2010 & dataset creators, data curators & Hybrid & Markup format \\ \hline
    \textit{Completeness of Datasets Documentation on ML/AI Repositories: An Empirical Investigation \cite{rondinaCompletenessDatasetsDocumentation2023}} & 2023 & dataset creators, data curators & Manual & Markup format \\ \hline
    \textit{Croissant: A Metadata Format for ML-Ready Datasets \cite{akhtarCroissantMetadataFormat2024}} & 2024 & dataset practitioners & Automated & Markup format \\ \hline
    \textit{CrowdWorkSheets: Accounting for individual and collective identities underlying crowdsourced dataset annotation \cite{diaz_crowdworksheets_2022}} & 2022 & dataset creators & Manual & Framework \\ \hline
    \textit{Data cards: Purposeful and transparent dataset documentation for responsible AI \cite{Data_Cards_Pushkarna_Zaldivar_Kjartansson_2022}} & 2022 & dataset creators, dataset users & Hybrid & Framework \\ \hline
    \textit{Data Readiness Report \cite{afzal_data_2021}} & 2021 & dataset practitioners, dataset creators & Manual & Framework \\ \hline
    \textit{Data Statements | Tech Policy Lab \cite{mcmillan-major_data_2023}} & 2023 & dataset creators, dataset users & Manual & Toolkit \\ \hline
    \textit{Data Statements for Natural Language Processing: Toward Mitigating System Bias and Enabling Better Science \cite{bender_data_2018}} & 2018 & dataset creators & Manual & Framework \\ \hline
    \textit{Data statements: From technical concept to community practice \cite{mcmillan-major_data_2024}} & 2024 & dataset creators & Manual & Schema \\ \hline
    \textit{Data-envelopes for cultural heritage: Going beyond datasheets \cite{luthra_data-envelopes_2024}} & 2024 & dataset creators & Manual & Framework \\ \hline
    \textit{DataDoc analyzer: A tool for analyzing the documentation of scientific datasets \cite{giner-miguelez_datadoc_2023}} & 2023 & dataset creators, dataset users & Automated & Application \\ \hline
    \textit{dataMaid: Your Assistant for Documenting Supervised Data Quality Screening in R \cite{petersenDataMaidYourAssistant2019}} & 2019 & dataset creators & Hybrid & Application \\ \hline
    \textit{Dataset construction challenges for digital forensics \cite{horsman_dataset_2021}} & 2021 & dataset creators, dataset users & Manual & Framework \\ \hline
    \textit{Datasheet for subjective and objective quality assessment datasets \cite{barman_datasheet_2023}} & 2023 & dataset creators, dataset users & Manual & Datasheet \\ \hline
    \textit{Datasheets for AI and medical datasets (DAIMS): a data validation and documentation framework before machine learning analysis in medical research \cite{marandi_datasheets_2025}} & 2025 & dataset creators, data experts & Hybrid & Datasheet \\ \hline
    \textit{Datasheets for datasets \cite{gebru_datasheets_2021}} & 2021 & dataset creators, dataset users & Manual & Datasheet \\ \hline
    \textit{Datasheets for datasets help ML engineers notice and understand ethical issues in training data \cite{boyd_datasheets_2021}} & 2021 & N/A & N/A & Study \\ \hline
    \textit{Datasheets for Energy Datasets: An Ethically-Minded Approach to Documentation \cite{heintz_datasheets_2023}} & 2023 & dataset creators & Manual & Datasheet \\ \hline
    \textit{Datasheets for Healthcare AI: A Framework for Transparency and Bias Mitigation \cite{siddik_datasheets_2025}} & 2025 & dataset creators, data experts & Automated & Framework \\ \hline
    \textit{DescribeML: a tool for describing machine learning datasets \cite{giner-miguelez_describeml_2022}} & 2022 & dataset creators & Hybrid & Application \\ \hline
    \textit{Ensuring Dataset Quality for Machine Learning Certification \cite{picardEnsuringDatasetQuality2020}} & 2020 & dataset creators, dataset users & Hybrid & Framework \\ \hline
    \textit{FactSheets: Increasing trust in AI services through supplier's declarations of conformity \cite{Fact_Sheets_Arnold_Bellamy_Hind_Houde_Mehta_Mojsilović_Nair_Ramamurthy_Olteanu_Piorkowski_2019}} & 2019 & dataset creators & Manual & Datasheet \\ \hline
    \textit{Goods: Organizing google's datasets \cite{halevy_goods_2016}} & 2016 & dataset creators, dataset users, dataset auditors & Hybrid & Application \\ \hline
    \textit{Healthsheet: Development of a Transparency Artifact for Health Datasets \cite{rostamzadehHealthsheetDevelopmentTransparency2022a}} & 2022 & dataset creators, data experts, data practitioners & Manual & Datasheet \\ \hline
    \textit{How to Automatically Document Data With the codebook Package to Facilitate Data Reuse \cite{arslan2019automatically}} & 2019 & dataset creators & Automated & Application \\ \hline
    \textit{Interactive Model Cards: A Human-Centered Approach to Model Documentation \cite{crisanInteractiveModelCards2022a}} & 2022 & N/A & N/A & study \\ \hline
    \textit{Language Dataset Documentation Design: Learning from Deaf and Indigenous Communities \cite{mcmillan-majorLanguageDatasetDocumentation2023}} & 2023 & dataset creators, dataset users & Manual & Toolkit \\ \hline
    \textit{Machine learning data practices through a data curation lens: An evaluation framework \cite{bhardwajMachineLearningData2024a}} & 2024 & dataset practitioners & Manual & Framework \\ \hline
    \textit{Method cards for prescriptive machine-learning transparency \cite{adkins_method_2022}} & 2022 & dataset creators & Manual & Datasheet \\ \hline
    \textit{MithraLabel: Flexible Dataset Nutritional Labels for Responsible Data Science \cite{sunMithraLabelFlexibleDataset2019}} & 2019 & dataset creators & Automated & Application \\ \hline
    \textit{Navigating Dataset Documentations in AI: A Large-Scale Analysis of Dataset Cards on Hugging Face \cite{Yang_Liang_Zou_2023}} & 2024 & N/A & N/A & Study \\ \hline
    \textit{Network report: a structured description for network datasets \cite{zheng_network_2022}} & 2022 & dataset creators, dataset users & Hybrid & Datasheet \\ \hline
    \textit{On Responsible Machine Learning Datasets with Fairness, Privacy, and Regulatory Norms \cite{mittal_responsible_2024}} & 2024 & dataset creators, data experts & Hybrid & Datasheet \\ \hline
    \textit{On the effectiveness of dataset watermarking \cite{atli_tekgul_effectiveness_2022}} & 2022 & dataset creators & Automated & Application \\ \hline
    \textit{Ontology-Supported AI Model and Dataset Management \cite{novacek_ontology-supported_2024}} & 2024 & dataset creators, dataset users & Hybrid & Application \\ \hline
    \textit{Open Datasheets: Machine-readable Documentation for Open Datasets and Responsible AI Assessments \cite{romanOpenDatasheetsMachinereadable2024a}} & 2024 & dataset creators & Hybrid & Application \\ \hline
    \textit{Prov-Dominoes: An approach for knowledge discovery from provenance data \cite{alencar_prov-dominoes_2024}} & 2024 & dataset creators & Hybrid & Application \\ \hline
    \textit{Reusable Templates and Guides For Documenting Datasets and Models for Natural Language Processing and Generation: A Case Study of the HuggingFace and GEM Data and Model Cards \cite{mcmillan-major_reusable_2021}} & 2021 & dataset creators, data organizations & Manual & Schema \\ \hline
    \textit{Right the docs: Characterising voice dataset documentation practices used in machine learning \cite{reid_right_2023}} & 2023 & N/A & N/A & Study \\ \hline
    \textit{Tackling algorithmic bias and promoting transparency in health datasets: the STANDING Together consensus recommendations \cite{alderman_tackling_2025}} & 2025 & dataset creators, data experts, dataset researchers & Manual & Framework \\ \hline
    \textit{Tackling Documentation Debt: A Survey on Algorithmic Fairness Datasets \cite{fabris_tackling_2022}} & 2022 & N/A & N/A & Study \\ \hline
    \textit{The ABC of Data: A Classifying Framework for Data Readiness \cite{castelijns_abc_2020}} & 2020 & dataset creators & Automated & Framework \\ \hline
    \textit{The CLeAR Documentation Framework for AI Transparency: Recommendations for Practitioners and Context for Policymakers \cite{The_CLeAR_Documentation_Framework_for_AI_Transparency_2024}} & 2024 & dataset practitioners & Manual & Framework \\ \hline
    \textit{The Dataset Nutrition Label (2nd Gen): Leveraging Context to Mitigate Harms in Artificial Intelligence \cite{chmielinskiDatasetNutritionLabel2022a}} & 2022 & dataset creators, data practitioners & Hybrid & Framework \\ \hline
    \textit{The Dataset Nutrition Label: A Framework To Drive Higher Data Quality Standards \cite{holland_dataset_2018}} & 2018 & dataset creators & Hybrid & Framework \\ \hline
    \textit{The State of Data Curation at NeurIPS: An Assessment of Dataset Development Practices in the Datasets and Benchmarks Track \cite{Bhardwaj_Gujral_Wu_Zogheib_Maharaj_Becker_2024}} & 2025 & N/A & N/A & Study \\ \hline
    \textit{Toward FAIR Semantic Publishing of Research Dataset Metadata in the Open Research Knowledge Graph \cite{ahmad_toward_2024}} & 2024 & dataset researchers & Hybrid & Markup format \\ \hline
    \textit{Towards accountability for machine learning datasets: Practices from software engineering and infrastructure \cite{hutchinson_towards_2021}} & 2021 & dataset creators & Manual & Framework \\ \hline
    \textit{Understanding machine learning practitioners' data documentation perceptions, needs, challenges, and desiderata \cite{heger_understanding_2022}} & 2022 & N/A & N/A & Study \\ \hline
    \textit{Using Large Language Models to Enrich the Documentation of Datasets for Machine Learning \cite{giner-miguelez_using_2024}} & 2024 & dataset practitioners & Automated & Application \\ 
    \arrayrulecolor{black}\bottomrule
    \noalign{\vskip 1em}
    \caption{\textcolor{black}{Items included in the review, along with the level of automation for each tool as described in Section~\ref{sec:methods:data-extraction}. Audiences are based on target audiences identified by authors in their contributions.}}
    \label{fig:table-corpus-definitions}
    \end{longtable}
\end{small}
\twocolumn

\end{document}